\documentclass{aa}  

\usepackage{graphicx}
\usepackage{txfonts}

\usepackage{hyperref}
\hypersetup{
    colorlinks=true,
    linkcolor=blue,
    filecolor=blue,      
    urlcolor=blue,
    citecolor=blue,
}
\usepackage{physics}
\usepackage{xfrac}
\usepackage{orcidlink}

\graphicspath{./figs/}

\newcommand*  {\Exp}[1]  {\mathrm{e}^{#1}}
\newcommand*  {\imag}    {\mathrm{i}}
\newcommand*  {\qed}     {\hbox{}\nobreak\hfill\quad\mbox{$\blacksquare$}}
\newcommand*  {\av}[1]   {\left\langle{#1}\right\rangle}
\newcommand*  {\psis}    {\psi_{\Sigma}}
\newcommand*  {\psip}    {\psi_{\Psi}}
\DeclareMathOperator{\sign}{sign}

\begin{document}

\title{The gravitational potential of spiral perturbations}
\subtitle{I. The 2D (razor-thin) case}

\author{Walter Dehnen\orcidlink{0000-0001-8669-2316}\thanks{E-mail: \href{mailto:walter.dehnen@uni-heidelberg.de}{walter.dehnen@uni-heidelberg.de}}}

\authorrunning{Walter Dehnen}

\institute{Astronomisches Rechen-Institut, Zentrum f{\"u}r Astronomie der Universit{\"a}t Heidelberg, M{\"o}nchhofstra\ss{}e~12-14, 69120, Heidelberg, Germany}

\date{Received June, 2025}

\abstract{
    I developed an efficient numerical method for obtaining the gravitational potential of razor-thin spiral perturbations and used it to assess the standard tight-winding approximation, which is found to be reasonably accurate for pitch angles $\alpha\lesssim20^\circ$. I derived the analytic potential of razor-thin logarithmic spirals with an arbitrary power-law amplitude. Approximating a spiral locally by one of these models provides a second-order tight-winding approximation that predicts the phase offset between the spiral potential and density, the resulting radially increasing pitch of the potential, and the nonlocal outward angular-momentum transport by gravitational torques. Beyond the inner and outer edge of a spiral with $m$ arms, its potential is not winding ($\alpha=90^\circ$), decays like $R^m$ and $R^{-1-m}$, respectively, and cannot be predicted by a local approximation.
}


\maketitle

\section{Introduction}
Spiral density perturbations are ubiquitous in late-type galaxies (by definition), and they are an important ingredient for the formation and evolution of these galaxies because the resulting non-axisymmetric gravitational forces alter the otherwise simple dynamics of the interstellar medium (ISM) and of the stars. The ISM is often strongly affected and forced to high densities, which triggers star formation, while the perturbation of the stellar orbits leads to radial migration and radial heating \citep{Sellwood2002}, which correspond to diffusion in the orbital actions $J_\phi$ and $J_R$, respectively.

Studies of these processes often require some model for the gravitational potential of the spiral perturbations, and in many situations, it is only necessary to model the dynamics in two dimensions without considering the vertical motions. Unfortunately, the only known gravitational potential for a flat (razor-thin) spiral density perturbation is that developed by \citeauthor{Kalnajs1971}' (\citeyear{Kalnajs1971}), which has an amplitude that declines with radius like $R^{-3/2}$ and is unsuitable as a model for galactic spirals. Instead, the tight-winding approximation (often also called `WKB approximation'\footnote{Called so in most papers and textbooks because the original development by \cite{LinShu1964} was similar to the (Jeffrey-)Wentzel-Kramers-Brillouin approximation of quantum mechanics. I abstain from using this term, because in the context of spiral perturbations some authors associate it with more than merely the approximation of the gravitational potential. \label{foot:WKB}}) allows reasonably accurate estimates for the gravitational potential of a spiral pattern with a pitch angle $\alpha\lesssim20^\circ$. 

Dynamical modellers typically employ simple models for the gravitational potential \citep[e.g.][]{Roberts1979, Contopoulos1986, Eilers2020, Hamilton2024} that are often based on this approximation. However, the actual density corresponding to these models has never been computed. Similarly, the actual potential generated by reasonably realistic spiral density perturbations is largely unknown.

The purpose of this study is to fill these gaps. To this end, I implement in Appendix~\ref{app:Poisson:numerics} an efficient numerical tool for finding the gravitational potential for any flat density and, vice versa, the density from the potential known in the equatorial plane only. This tool is then applied to assess the accuracy of the tight-winding approximations, including a more accurate novel version, for logarithmic spirals with amplitude that declines with radius either as a power law or exponentially (like the surface density of disc galaxies) with and without radial truncation. I also obtain the actual density of models for spiral potentials that have been used in previous studies, and find that the density sometimes behaves in unexpected ways. For example, the pitch angle can deviate significantly from the target.

Finally, I consider the gravitational torques of a spiral perturbation and derive an approximate expression for their local mean at each radius. For trailing spirals, this is always negative at small and positive at large radii, implying an outward angular-momentum transport \citep{Zhang1996}, in addition to the advective angular momentum transport by diffusion driven by the standard deviation of the torques.

I begin the presentation in Section~\ref{sec:spiral} by introducing my notations and giving a brief overview over known results, including the tight-winding approximation. In Section~\ref{sec:spiral:scale-invariant}, I extend \citeauthor{Kalnajs1971}' (\citeyear{Kalnajs1971}) potential-density pairs to logarithmic spirals whose density amplitude is an arbitrary power law in radius. I also develop a novel more accurate tight-winding approximation. Section~\ref{sec:torque} considers the non-local gravitational torque induced by a spiral perturbation. In Section~\ref{sec:pot}, I consider various models for spiral perturbations and compare their numerically computed properties to the predictions of the tight-winding approximations. Finally, Sections~\ref{sec:discuss} and~\ref{sec:conclude} discuss my findings and present my conclusion.

\section{Spiral density perturbations}
\label{sec:spiral}
This section introduces basic concepts and notations and summarises the important tight-winding approximation, but presents no novel results.

\subsection{Poisson's equation for razor-thin systems}
\label{sec:notation}
This study exclusively considers the situation of razor-thin mass distributions with some surface density $\Sigma(\vec{R})$, where $\vec{R}\equiv(x,y)$, and the spatial density $\rho(\vec{r}) = \Sigma(\vec{R})\,\delta(z)$, where $\vec{r}\equiv(x,y,z) = (\vec{R},z)$. The gravitational potential
\begin{align}
    \label{eq:Phi:from:dens}
    \Phi(\vec{r}) = -G\iiint\frac{\rho(\vec{r}')\,\dd^3\vec{r}'}{|\vec{r}-\vec{r}'|}
\end{align}
satisfies the Poisson equation $\laplacian\Phi(\vec{r}) = 4\pi G\,\rho(\vec{r})$, which implies 
\begin{align}
    \label{eq:Surf:from:Phi}
    \Sigma(\vec{R}) = \frac1{2\pi G}\lim_{z\to0^+}\pdv{\Phi}{z}.
\end{align}
However, often only the potential in the equatorial plane $\Psi(\vec{R})\equiv\Phi(\vec{R},z=0)$ is required or even known, when Eq.~\eqref{eq:Phi:from:dens} becomes
\begin{align}
    \label{eq:Psi:from:Surf}
    \Psi(\vec{R}) = 
    -G\iint\frac{\Sigma(\vec{R}')\,\mathrm{d}^2\vec{R}'}{|\vec{R}-\vec{R}'|}.
\end{align}
The inverse operation, obtaining $\Sigma(\vec{R})$ from $\Psi(\vec{R})$, is not as straightforward as for three-dimensional models because Eq.~\eqref{eq:Surf:from:Phi} cannot be used. Nonetheless, a useful relation can be inferred from the fact that a razor-thin plane density wave $\Sigma\propto\Exp{\imag\vec{k}\cdot\vec{R}}$ has the potential $\Phi(\vec{r})=\Psi(\vec{R})\Exp{-k|z|}$, where $k=|\vec{k}|$ and
\begin{align}
    \label{eq:Psi:wave}
    \Psi(\vec{R})=-2\pi G\Sigma(\vec{R})/k.
\end{align}
It follows that in Fourier space, Eq.~\eqref{eq:Psi:from:Surf} becomes $\hat{\Psi}=-2\pi G\hat{\Sigma}/k$ or $2\pi G\hat{\Sigma}=-k\hat\Psi=\widehat{\laplacian\Psi}/k$, where a hat denotes the Fourier transform. Thus, $-\laplacian\Psi$ relates to $2\pi G\Sigma$ in exactly the same way as $2\pi G\Sigma$ relates to $\Psi$, and therefore, 
\begin{align}
    \label{EQ:SURF:FROM:PSI}
    \Sigma(\vec{R}) = \frac1{4\pi^2G} \iint\frac{\laplacian\Psi(\vec{R}')\,\dd^2\vec{R}'}{|\vec{R}-\vec{R}'|}
\end{align}
\citep[H.~Dejonghe in][problem 2.22; see Appendix~\ref{app:proof} for a proof without Fourier transform]{BT2008}. Thus, obtaining $\Sigma(\vec{R})$ from $\Psi(\vec{R})$ requires a similar effort as the inverse relation and can be achieved with the same numerical or analytic tools. Appendix~\ref{app:Poisson:numerics} details the numerical tool I used in this study.

\subsection{Spiral patterns}
\label{sec:spiral:patters}
The density and potential of the disc can be uniquely decomposed into azimuthal Fourier components of the general form
\begin{subequations}
    \label{eqs:spiral}
\begin{align}
    \label{eq:spiral:Surf}
    \Sigma(\vec{R}) &= \phantom{-} S_{\!m}(R)  \, \Exp{\imag m [\phi - \psis(R)]},\\
    \label{eq:spiral:Psi}
    \Psi(\vec{R}) &= -P_{\!m}(R) \, \Exp{\imag m [\phi - \psip(R)]},
\end{align}
\end{subequations}
with amplitudes $P_{\!m}$, $S_{\!m}\ge0$ and phases $\psip$, $\psis$, which are real-valued functions of the radius $R=|\vec{R}|$. As usual, the physical potential and density are given by the real parts of Eqs.~\eqref{eqs:spiral}, and we can assume $m\ge0$ without loss of generality. For a spiral pattern, the phase is a monotonously increasing (or decreasing) function of radius. The pitch angle $\alpha>0$ between the wave crest at $\phi=\psi(R)$ and concentric circles relates to the phase via
\begin{align}
    \label{eq:pitch:def}
    \lambda\equiv\dv{\psi}{\ln R} = \frac\sigma{\tan\alpha},
\end{align}
where $\sigma=\pm1$ gives the sense in which the spiral twists, such that $\alpha>0$. Galactic spirals are trailing, implying $\sigma=-\sign(\bar{v}_\phi)$, where $\bar{v}_\phi$ is the mean azimuthal (rotational) velocity. The pitch angles of galactic spirals tend to vary with radius \citep{Kennicutt1981, Savchenko2013, Chugunov2025}, but a widespread model is that of a constant pitch angle, which is obtained for
\begin{align}
    \label{eq:psi:log}
    \psi(R) = \lambda \ln(R/R_0)
\end{align}
and referred to as a `logarithmic spiral'. In this case, the phase factors in Eqs.~\eqref{eqs:spiral} become $\Exp{\imag m[\phi - \lambda\ln(R/R_0)]}$, which is a plane wave in both $\phi$ and $\ln R$.

Except for models with power-law amplitudes (see Section~\ref{sec:spiral:scale-invariant}), no analytic solutions to Eq.~\eqref{eq:Psi:from:Surf} are known for spirals of the general form~\eqref{eq:spiral:Surf}. Instead, the following approximation is often employed.

\subsection[]{The tight-winding (or WKB) approximation}
\label{sec:spiral:WKB}
Each annulus of a disc with, say, an $m=2$ perturbation generates a quadrupolar gravitational field. For a bar-like perturbation, these quadrupoles have the same orientation, and their contributions to the field at some position accumulate. For a spiral, on the other hand, the orientations vary with radius, and most of the contributions from distant annuli to the local field cancel (destructive interference), such that it is dominated by the local pattern. To estimate the field from the local pattern, the curvature of the spiral can be ignored and it can be approximated as a plane wave. In analogy to the relation~\eqref{eq:Psi:wave}, this gives \citep{Toomre1964, LinShu1964}
\begin{align}
    \label{eq:WKB}
    \Psi(\vec{R}) &\approx -\frac{2\pi G}{k(R)}\Sigma(\vec{R}),
\end{align}
where $k(R)$ is a locally appropriate horizontal wavenumber. Following \cite{LinShu1964}, $k(R)$ is traditionally taken to be just the radial wavenumber 
\begin{subequations}
    \label{eqs:WKB:k:1}
\begin{align}
    \label{eq:WKB:k:tan}
    k(R) = |k_R| = \frac{m\qty|\lambda|}{R} = \frac{m}{R\tan\alpha}.
\end{align}
However, the analogy to plane waves suggests the absolute value of the horizontal wavevector, which also includes the azimuthal wavenumber $k_\phi=m/R$, i.e.
\begin{align}
    \label{eq:WKB:k:sin}
    k(R) = \sqrt{k_R^2+k_\phi^2} = \frac{m\sqrt{\lambda^2+1}}R = \frac{m}{R\sin\alpha}.
\end{align}
\end{subequations}
Thus, the potential is weaker for more arms (larger $m$) and more tightly wound spirals (smaller $\alpha$). Apparently, Eq.~\eqref{eq:WKB:k:sin} has not been used explicitly, but is implicit in shearing-sheet approaches for modelling spiral structure \citep[e.g.][Eq.~11]{JulianToomre1966}. Both Eqs.~\eqref{eqs:WKB:k:1} are first-order accurate in the sense that the fractional error made in Eq.~\eqref{eq:WKB} is $\order*{\sin\alpha}$ and the total error $\order*{\sin^2\!\alpha}$, but Eq.~\eqref{eq:WKB:k:sin} is significantly better than Eq.~\eqref{eq:WKB:k:tan}, as demonstrated in Figs.~\ref{fig:scale-invariant:WKB} and \ref{fig:exp:Pot:WKB+SFA}.

In general, the phases $\psip$ and $\psis$ of the potential and density of a spiral perturbation differ, but the first-order approximations~\eqref{eqs:WKB:k:1} give $\psip=\psis$ and hence do not predict the phase offset $\delta\psi\equiv\psip-\psis$. \citeauthor{Shu1970}'s (\citeyear{Shu1970}) second-order approximation 
\begin{align}
    \label{eq:WKB:Shu}
    \Sigma(\vec{R}) = -\frac{\imag\sigma}{2\pi G\sqrt{R}} \pdv{\sqrt{R}\Psi}{R} \;\times\;\left[1+ O(\sin^2\!\alpha)\right],
\end{align}
predicts $\delta\psi$, but does not provide $\Psi(\vec{R})$ for a given $\Sigma(\vec{R})$.

\begin{figure}
    \includegraphics[width=\columnwidth]{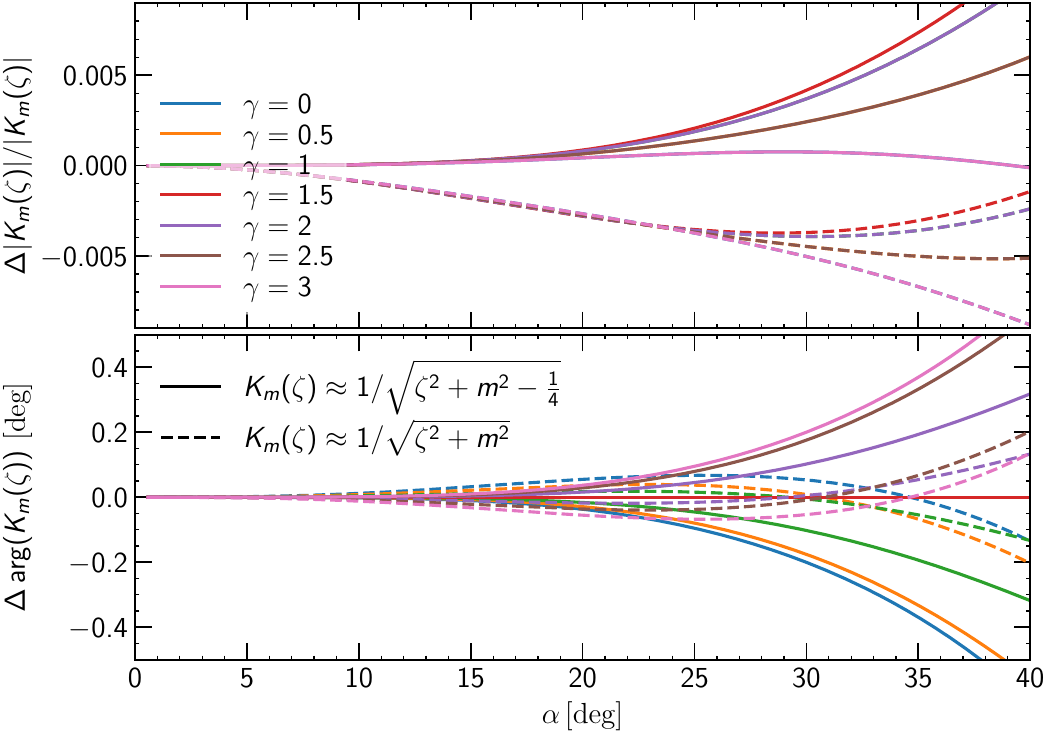}
    \vspace*{-6mm}
    \caption{
    Assessing the accuracy of the approximation~\eqref{eq:K:asymp} for $m=2$ (for $m>2$ the accuracy is better). The real (imaginary) parts are even (odd) functions of $\gamma-\sfrac32$ (hence the relations for $\gamma<\sfrac32$ are hidden in the top panel). I also show (dashed) the relation given by \cite{Kalnajs1971}.}
    \label{fig:scale-invariant:App}
\end{figure}

\section{Scale-invariant spirals}
\label{sec:spiral:scale-invariant}
A simple situation is that of a logarithmic spiral whose density amplitude follows a power law in radius,
\begin{align}
    \label{eq:Sigma:power-law}
    \Sigma(R,\phi) &= S_{\!m,0}\, (R/R_0)^{-\gamma} \Exp{\imag m[\phi-\lambda\ln(R/R_0)]},
\end{align}
such that the total $R$ dependence, $\Sigma\propto R^{-\gamma-\imag m\lambda}$, is still a power law and hence scale-invariant. The gravitational potential of these models is (see Appendix~\ref{app:K} for a derivation)
\begin{align}
    \label{eq:scale-invariant:Phi}
    \Phi(\vec{r}) &= -2\pi G\,f_m(|z|/r)\,r\,\Sigma(r,\phi),
\end{align}
with spherical radius $r=|\vec{r}|$ and
\begin{align}
    \label{eq:Km}
    f_m(x) &= -\qty(\dd P^{\,|m|}_{-1/2-\imag\zeta}/\dd x)^{-1}_{x=0}\, P^{\,|m|}_{-1/2-\imag\zeta}(x),
\end{align}
where 
\begin{align}
    \label{eq:zeta}
    \zeta &\equiv m\lambda-\imag(\gamma-\sfrac32)
\end{align}
and $P_\nu^{\,m}(x)$ denotes the associated Legendre function of the first kind. The function $f_m$ describes the run of $\Phi(\vec{r})$ with latitude and at $x\to1$ declines like $(1-x)^{m/2}$. 

This study focusses on the potential in the $z=0$ plane,
\begin{align}
    \label{eq:Psi:From:Surf:scale-invariant}
    \Psi(\vec{R})= -2\pi G\,K_m(\zeta)\,R\,\Sigma(\vec{R}),
\end{align}
where
\begin{align}
    \label{eq:Km0}
    K_m(\zeta) \, &\equiv f_m(0) =\frac12\frac{\Gamma\Big(\frac14+\frac{m}2+\imag\frac\zeta2\Big)\,\Gamma\Big(\frac14+\frac{m}2-\imag\frac\zeta2\Big)}{\Gamma\Big(\frac34+\frac{m}2+\imag\frac\zeta2\Big)\,\Gamma\Big(\frac34+\frac{m}2-\imag\frac\zeta2\Big)},
\end{align}
which satisfies $K_m(-\zeta)=K_m(\zeta)$, $K_m(\zeta^\ast)=K_m^\ast(\zeta)$ (where a star denotes complex conjugation), and
\begin{align}
    \label{eq:K:prop:1}
    K_m(\zeta)\,K_{m+1}(\zeta)
    &= \left[\zeta^2+\big(m+\tfrac12\big)^2\right]^{-1}, \\
    K_m(\zeta)\,K_m(\zeta+\imag)
    &= \left[\big(\zeta+\tfrac\imag2\big)^2+m^2\right]^{-1}.
\end{align}
Interpolating these relations gives the asymptotic behaviour
\begin{align}
    \label{eq:K:asymp}
    K_m(\zeta) &\approx \left[\zeta^2+m^2-\tfrac14\right]^{-1/2},
\end{align}
which is an excellent approximation even for pitch angles as large as $40^\circ$, as Fig.~\ref{fig:scale-invariant:App} demonstrates.

For these scale-invariant spirals, the potential $\Psi$ and density $\Sigma$ are proportional to the same phase factor $\Exp{\imag m[\phi - \lambda\ln(R/R_0)]}$, such that they have the same constant pitch angle $\alpha$. For $\gamma=\sfrac32$, $\zeta$ is real-valued, and hence, so are $f_m(x)$ and $K_m(\zeta)$, such that $\Psi$ and $\Sigma$ even have the same phase $\psi(R)$. In this case, $S_{\!m}\propto R^{-3/2}$ and $P_{\!m}\propto R^{-1/2}$, and we obtain the potential-density pairs of \cite{Kalnajs1971}, who for real-valued $\zeta$ gave $\Psi(\vec{R})$ with Eqs.~(\ref{eq:Km0},\ref{eq:K:prop:1}) and an inferior version of Eq.~\eqref{eq:K:asymp}. 

\begin{figure}
\begin{center}
	\includegraphics[width=0.97\columnwidth]{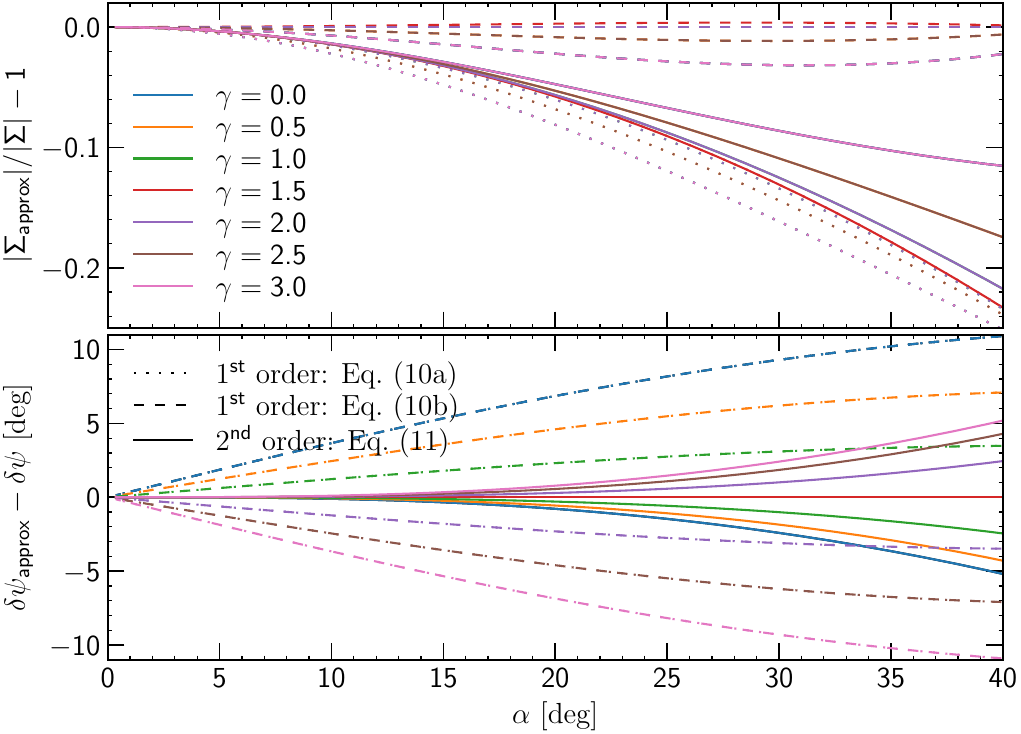}
    \vspace*{-2mm}
    \caption{
    Assessing the tight-winding (WKB) approximation for scale-invariant spirals. The relative error of the density amplitude and the error in the phase offset $\delta\psi\equiv\psip-\psis$ are plotted vs.\ pitch angle $\alpha$ for $m=2$ and various values of the exponent $\gamma$ (for these scale-invariant models, the errors are the same at each radius). Since the first-order approximations give $\delta\psi_{\mathrm{approx}}=0$, their error reflects the actual phase offset of the models.}
    \label{fig:scale-invariant:WKB}
\end{center}
\vspace*{-8mm}
\end{figure}

For $\gamma\neq\sfrac32$, however, $K_m(\zeta)$ is complex-valued and causes a constant phase offset $\delta\psi=-\arg(K_m(\zeta))/m$ of $\Psi$ with respect to $\Sigma$ (see also Fig.~\ref{fig:scale-invariant:WKB}), such that for a trailing spiral, the potential lags the density for $\gamma<\sfrac32$ and leads it for  $\gamma>\sfrac32$ \citep{Zhang1996}.

As shown in the next section, this constant offset would imply a non-vanishing total torque in contradiction to Noether's theorem. However, when the total torque is expressed as an integral over pairs of annuli torquing each other, it is found to be zero. This contradiction arises because of the unphysical nature of these models at $R\to0$ (see Appendix~\ref{app:torque} for details).

\subsection{Comparison with the tight-winding approximation}
\label{sec:WKB:cmp}
The scale-invariant spirals have no inner or outer edge, such that there is a local spiral field everywhere. This makes these models the ideal situation for the tight-winding approximation. Fig.~\ref{fig:scale-invariant:WKB} plots the errors in amplitude and phase made by the first-~\eqref{eqs:WKB:k:1} and second-order approximation~\eqref{eq:WKB:Shu}, respectively. The traditional tight-winding approximations~\eqref{eq:WKB:k:tan} and~\eqref{eq:WKB:Shu} underestimate the density or, conversely, overestimate the potential and forces, but the first-order approximation~\eqref{eq:WKB:k:sin} only makes very small errors in amplitude. The order of the approximations is only obvious in the phase error (a peculiarity of these power-law models), where the first-order methods predict $\delta\psi=0$, while the phase error of the second-order approximation is quadratic in the pitch angle, as expected.

For $\gamma=\sfrac32$ (red; \citeauthor{Kalnajs1971}' \citeyear{Kalnajs1971} spirals), $\psip=\psis$ and the approximation using Eq.~\eqref{eq:WKB:k:sin} is identical to Eq.~\eqref{eq:K:asymp} and hence extremely accurate. Away from the $z=0$ plane, however, the tight-winding approximation deteriorates, as it predicts $\psip(R,z)=\psis(R)$, while actually $\psip(R,z)=\psis(\sqrt{R^2+z^2})$, i.e.\ the phase is in fact constant on spheres, not on cylinders. A more detailed assessment of the potential and its approximation away from the plane is beyond the scope of this study.

\subsection{An improved tight-winding approximation}
\label{sec:spiral:approx:scale-invariant}
In the traditional tight-winding approximation, a spiral is locally approximated as plane wave, i.e.\ assumed to have constant amplitude and pitch angle $\alpha\propto R$. The amplitude might instead be assumed to decline locally like a power law and the pitch angle be constant. In other words, the spiral is approximated locally as a scale-invariant spiral. When its potential is in turn approximated via Eq.~\eqref{eq:K:asymp}, it can be expressed in the standard form~\eqref{eq:WKB} using the complex-valued wavenumber
\begin{align}
    \label{eq:WKB:k:SFA}
    k(R) = \frac1R \sqrt{{\zeta^2(R)+m^2-\tfrac14}},
\end{align}
where the principal value of the square root is taken and
\begin{align}
    \label{eq:zeta(R)}
    \zeta(R) &= \imag \left[\frac32+\frac{R}{\Sigma}\pdv{\Sigma}{R}\right]
     = m\lambda(R) - \imag \left(\gamma(R)-\tfrac32\right),
\end{align}
with $\lambda(R)$ as defined in Eq.~\eqref{eq:pitch:def} and $\gamma(R)\equiv-\dd\ln S_{\!m}/\dd\ln R$.

It is also possible to set $\zeta(R)=\imag[\sfrac12 + (R/\Psi)(\partial\Psi/\partial R)]$, which allows the approximation of $\Sigma(\vec{R})$ given $\Psi(\vec{R})$, and, when also approximating $\sqrt{\zeta^2+m^2-\sfrac14}\approx\sigma\zeta$, gives the second-order approximation~\eqref{eq:WKB:Shu}. A somewhat more simply computable and manipulable form than Eq.~\eqref{eq:WKB:k:SFA} is provided by the Taylor expansion,
\begin{align}
    \label{eq:SFA:expn}
    \frac1{k(R)} \approx
    \frac{R}{\sqrt{\zeta^2+m^2}}
    &= \frac{R}{m\sqrt{\lambda^2+1}} \left[1 +  
    \frac{\imag\lambda\epsilon}{\sqrt{\lambda^2+1}}  + \order*{\epsilon^2} \right],
\end{align}
where $\epsilon\equiv(\gamma-\sfrac32)/m\sqrt{\lambda^2+1}=\sin\alpha\,(\gamma-\sfrac32)/m$. When this is truncated at zeroth order in $\epsilon$ (or for $\gamma=\sfrac32$ when $\epsilon=0$), this collapses to the first-order approximation~\eqref{eq:WKB:k:sin}. These relations suggest that the novel approximation~\eqref{eq:WKB:k:SFA} is at least second-order accurate. In terms of the pitch angle $\alpha$ (instead of $\lambda$), the approximation~\eqref{eq:WKB} resulting from the expansion~\eqref{eq:SFA:expn} is conveniently expressed as
\begin{align}
    \label{eq:WKB:novel}
    \Psi(\vec{R}) &\approx -2\pi G \frac{\sin\alpha}{m}\left[1+\imag\sigma\frac{\sin2\alpha}{2m}\left(\gamma-\tfrac32\right)\right]R\,\Sigma(\vec{R}).
\end{align}
Section~\ref{sec:pot} assesses this together with the first-order approximations for various spiral models.

\section{The gravitational torque of spiral perturbations}
\label{sec:torque}
A spiral perturbation exerts a gravitational torque per unit mass of $-\partial\Phi/\partial\phi$. This induces changes in the stellar angular momenta, which cause radial migration \citep{Sellwood2002} and are particularly pronounced for stars that co-rotate with the spiral pattern (and for stars on the Lindblad resonances) because they experience coherent torquing, while the torques largely average away for other stars. The distribution of angular-momentum changes $\delta J_\phi$ induced by these torques per unit time is characterised by its first two moments $\langle\delta J_\phi\rangle$ and $\langle(\delta J_\phi)^2\rangle$. The second moment determines the local diffusion of $J_\phi$, which causes advective angular-momentum transport.

The first moment, $\langle\delta J_\phi\rangle$, describes non-local angular-momentum transport and is caused by the average torque at fixed $J_\phi$. While the azimuthal average of $-\partial\Phi/\partial\phi$ vanishes, that of the torque density $\tau\equiv-\Sigma(\partial\Psi/\partial\phi)$ does not in general vanish because the distribution of stars is not axially symmetric owing to the spiral pattern itself. For potential and density perturbations of the form~\eqref{eqs:spiral}, the azimuthal average of $\tau$ is \citep{Zhang2007}
\begin{align}
    \label{eq:ave:tau}
    \av{\tau}_\phi(R) = \tfrac12 m\, S_{\!m}(R)P_{\!m}(R)\sin\big(m\,\delta\psi(R)\big),
    \;\;\;\delta\psi \equiv \psip - \psis
\end{align}
if $m_\Psi=m_\Sigma=m$ and $\av{\tau}_\phi=0$ otherwise. Thus, if the potential and density are out of phase at radius $R$, a spiral perturbation exerts a net torque over the annulus. For a trailing spiral, if $\sigma\,\delta\psi<0$ the potential leads the density, and stars gain angular momentum on average. Conversely, if $\sigma\,\delta\psi>0$, the potential trails the density, and stars lose angular momentum on average.

According to Noether's theorem, the total angular momentum is conserved. The total torque over the whole disc, $T{\,=\,}2\pi \int_0^\infty\av{\tau}_\phi R\,\dd R$, therefore vanishes (as shown explicitly in Appendix~\ref{app:torque}), and these gains and losses at different radii balance, but constitute a non-local transport of angular momentum from regions where $\sigma\delta\psi>0$ to those where $\sigma\delta\psi<0$. 

The azimuthally averaged torque density $\av{\tau}_\phi$ can be computed from the numerically computed potential for any spiral density perturbation, but a simple way to approximate it would be insightful.

\subsection{Torque in the tight-winding approximation}
\label{sec:torque:approx:WKB}
For a density component of the form~\eqref{eq:spiral:Surf}, the approximation~\eqref{eq:WKB} gives for the azimuthally averaged torque density
\begin{align}
    \label{eq:tau:WKB}
    \av{\tau}_\phi &\approx - m \pi G S_{\!m}^{\!2} \Im\{k^{-1}\}.
\end{align}
The first-order approximations~\eqref{eqs:WKB:k:1} have real-valued $k(R)$ and hence give $\av\tau_\phi=0$. This is not surprising: As the product of the first-order quantities $P_{\!m}$ and $\sin\delta\psi$, the torque is a second-order quantity and cannot be predicted by first-order approximations. For the novel approximation~\eqref{eq:WKB:novel}, we find 
\begin{align}
    \label{eq:tau:novel}
    \av{\tau}_\phi
    &\approx -\frac{\sigma\pi}{2m}  G R\,S_{\!m}^{\!2} \sin\alpha\,\sin2\alpha\,\left(\gamma-\tfrac32\right).
\end{align}

\begin{figure*}
	\includegraphics[width=0.995\columnwidth]{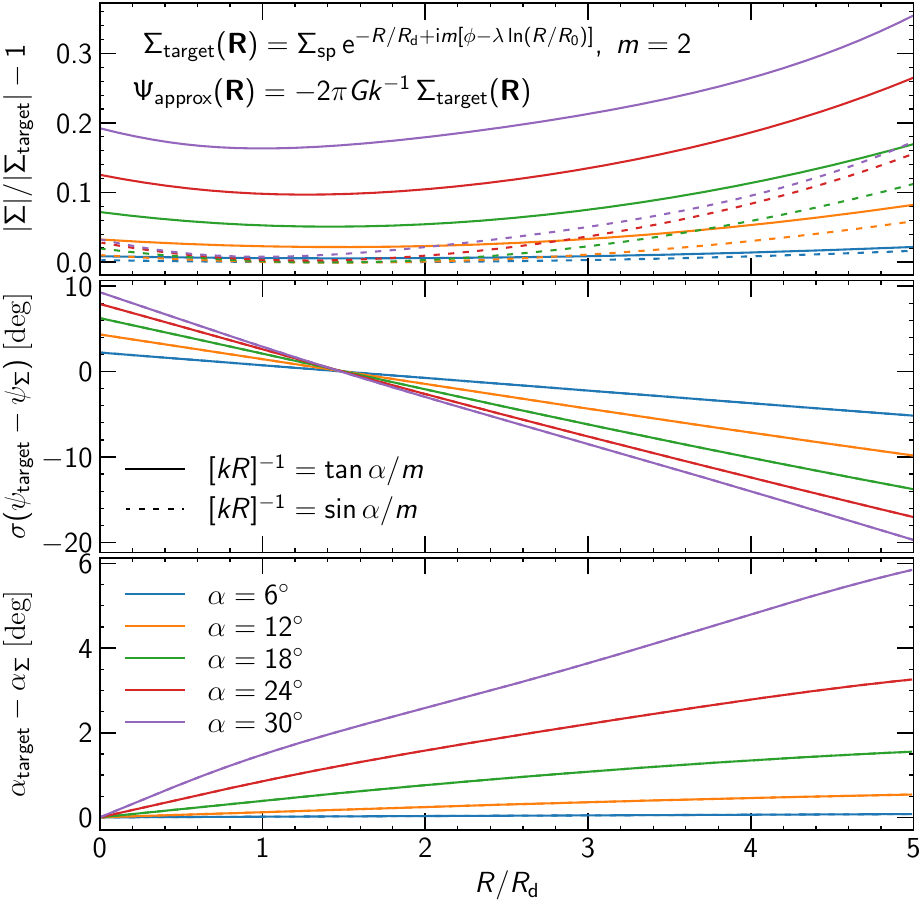} \hfill
	\includegraphics[width=0.995\columnwidth]{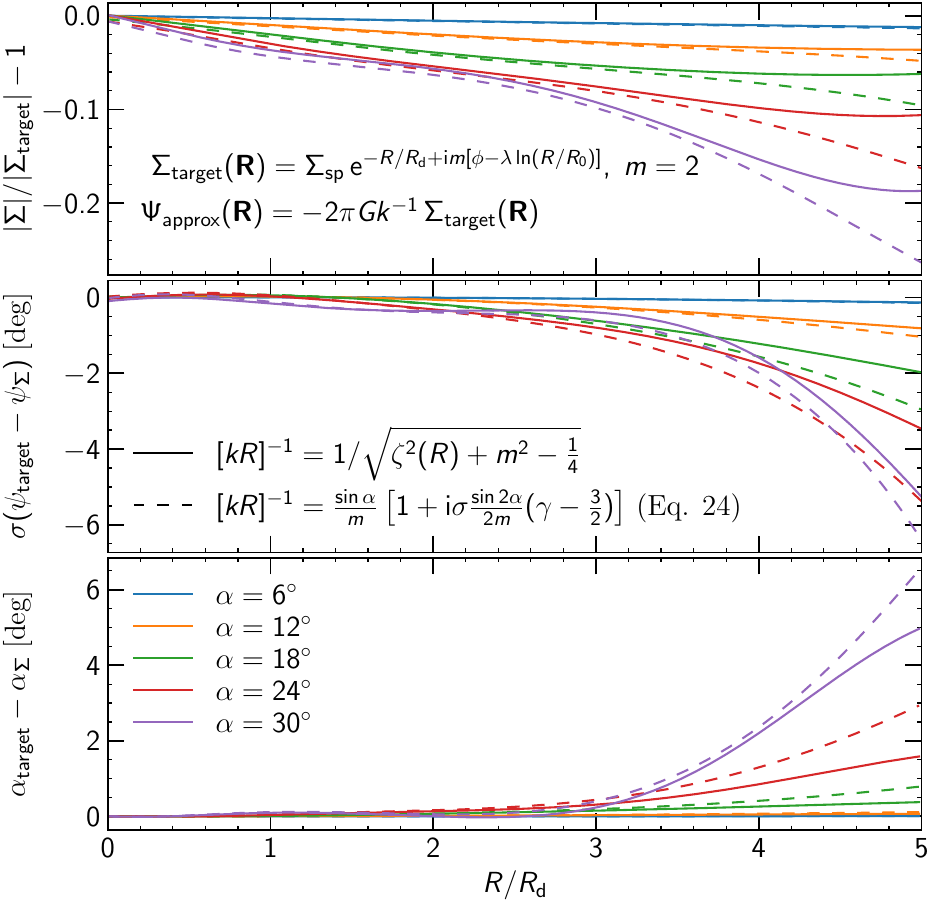}
    \vspace*{-2mm}
    \caption{Assessing the ability of the first-order (\textbf{left}) and our novel (\textbf{right}) tight-winding approximations to provide a gravitational potential for the target density~\eqref{eq:Sigma:exp:target} of a logarithmic spiral with pitch angle $\alpha$ and exponentially declining amplitude. From the density that is actually generated by the approximate potential, I plot the radial profiles of the relative amplitude error (\textbf{top}) and the errors in phase (\textbf{middle}) and pitch angle (\textbf{bottom}), which for the first-order methods are also the respective offsets between the potential and density because to first order, $\psip=\psi_{\mathrm{target}}$.}
    \label{fig:exp:Pot:WKB+SFA}
    \vspace*{-3mm}
\end{figure*}

\subsection{Torque from the potential}
\label{sec:torque:approx:pot}
Instead of estimating $\av{\tau}_\phi$ from the target density and its second-order approximate potential as above, I now compute it for a first-order approximate potential and its (unknown) actual density using the exact expression \citep[][Eq.~6.18]{BT2008}
\begin{align}
    \label{eq:T(<R)}
    T(R) = -\frac{R}{4\pi G} \int_0^{2\pi}\dd\phi\int_{-\infty}^\infty\dd z\eval{\pdv{\Phi}{R}\pdv{\Phi}{\phi}}_R
\end{align}
for the cumulative disc torque $T(R)\equiv2\pi\int_0^R \dd R' R'\av{\tau}_\phi(R')$. In order to apply this, we must evaluate the integral over $z$. For the first-order tight-winding approximation based on plane waves, the natural approximation is $\Phi(\vec{r}) = \Psi(\vec{R}) \Exp{-k(R)|z|}$, when the integral over $z$ gives a factor $1/k(R)$, and we obtain after inserting Eqs.~\eqref{eq:spiral:Surf} and~\eqref{eq:WKB}, taking the derivatives, integrating over $\phi$ (and recalling that only the real part of $\Psi$ has meaning)
\begin{align}
    \label{eq:T(<R):approx}
    T(R) &\approx \frac{\sigma\pi Gm^2}{\tan\alpha}\frac{S_{\!m}^{\!2}(R)}{k^3(R)\,}.
\end{align}
The azimuthally averaged torque density can now be estimated as $\av\tau_\phi=(\dd T/\dd R)/(2\pi R)$, which for the approximation~\eqref{eq:WKB:k:sin} again gives Eq.~\eqref{eq:tau:novel}. As $T(R)$ in Eq.~\eqref{eq:T(<R):approx} vanishes for $R\to\infty$ (provided $S_{\!m}$ decays faster than $R^{-3/2}$ in this limit), this also implies that the estimate~\eqref{eq:tau:novel} obtains zero total torque exactly.

\section{Application: The potential of spirals}
\label{sec:pot}
I now consider various models for spiral perturbations and study the associated gravitational potentials, but also assess the ability of the tight-winding approximations to predict them.

\subsection{Spirals with an exponential amplitude profile}
\label{sec:exp}
The unperturbed ($m=0$) surface density of disc galaxies generally follows an exponential decline,
\begin{align}
    \label{eq:disc:exp}
    \Sigma_0(R)=\Sigma_{\mathrm{c}} \Exp{-R/R_{\mathrm{d}}}.
\end{align}
A simple and natural model for a spiral is one with a constant relative amplitude $A_m\equiv|\Sigma_m|/\Sigma_0$ (Section~\ref{sec:taper:real} considers $A_m$ to depend on radius). Logarithmic spirals with this amplitude have the density
\begin{align}
    \label{eq:Sigma:exp:target}
    \Sigma(\vec{R}) = \Sigma_{\mathrm{sp}} \Exp{-R/R_{\mathrm{d}}+\imag m[\phi-\lambda\ln(R/R_0)]},
\end{align}
where $\Sigma_{\mathrm{sp}}=A_m \Sigma_{\mathrm{c}}$. I now consider various tight-winding approximations for the gravitational potential as well as that computed numerically.

\subsubsection{Assessing the tight-winding approximations}
A common model is
\begin{align}
    \label{eq:Psi:exp:simple}
    \Psi(\vec{R}) = -A_\Psi R \Exp{-R/R_{\mathrm{d}}+\imag m[\phi-\lambda\ln(R/R_0)]},
\end{align}
with some (often unspecified) constant $A_\Psi$, which corresponds to the first-order approximation~\eqref{eq:WKB} with Eq.~\eqref{eq:WKB:k:tan} if
\label{eqs:WKB:exp}
\begin{align}
    \label{eq:WKB:exp:1}
    A_\Psi = 2\pi G\Sigma_{\mathrm{sp}} m^{-1}\tan\alpha
\end{align}
\citep[][and several subsequent studies]{Contopoulos1986} or Eq.~\eqref{eq:WKB:k:sin} if $A_\Psi = 2\pi G\Sigma_{\mathrm{sp}} m^{-1}\sin\alpha$. For constant and real-valued $A_\Psi$, the potential has the same phase as the target density~\eqref{eq:Sigma:exp:target}, such that $\alpha_\Psi=\alpha$. My novel approximation can be expressed in this form by setting $A_\Psi$ to a complex-valued function of radius. For five values of $\alpha$, I compute numerically (via Eq.~\ref{EQ:SURF:FROM:PSI} and the gravity solver of Appendix~\ref{app:Poisson:numerics}) the surface densities $\Sigma(\vec{R})$ corresponding to these potential approximations and compare in Fig.~\ref{fig:exp:Pot:WKB+SFA} their properties to those of the target~\eqref{eq:Sigma:exp:target}.

The first-order method using the non-standard $k(R)$ in Eq.~\eqref{eq:WKB:k:sin} approximates the amplitude significantly better than the standard form~\eqref{eq:WKB:k:tan}. However, the two first-order methods give the same phase for the density as for the potential, and hence, an incorrect phase offset and pitch angle for the density, while the novel approximation gives an excellent match for $R\lesssim3R_{\mathrm{d}}$ or $\alpha\lesssim20$ and still a reasonably accurate density phase at higher values. The full version of my novel approximation (Eq.~\ref{eq:WKB:k:SFA}) is somewhat better than its reduced form (Eq.~\ref{eq:SFA:expn} or~\ref{eq:WKB:novel}), but the difference is only significant in the regime where neither is very good any more.

The failure of all the tight-winding approximations at larger radii is not surprising because owing to the exponentially declining amplitude, the local field (on which these approximations are based) becomes ever less relevant compared to the far field from the inner parts (ignored by these approximations).

\begin{figure}
\begin{center}
	\includegraphics[width=0.985\columnwidth]{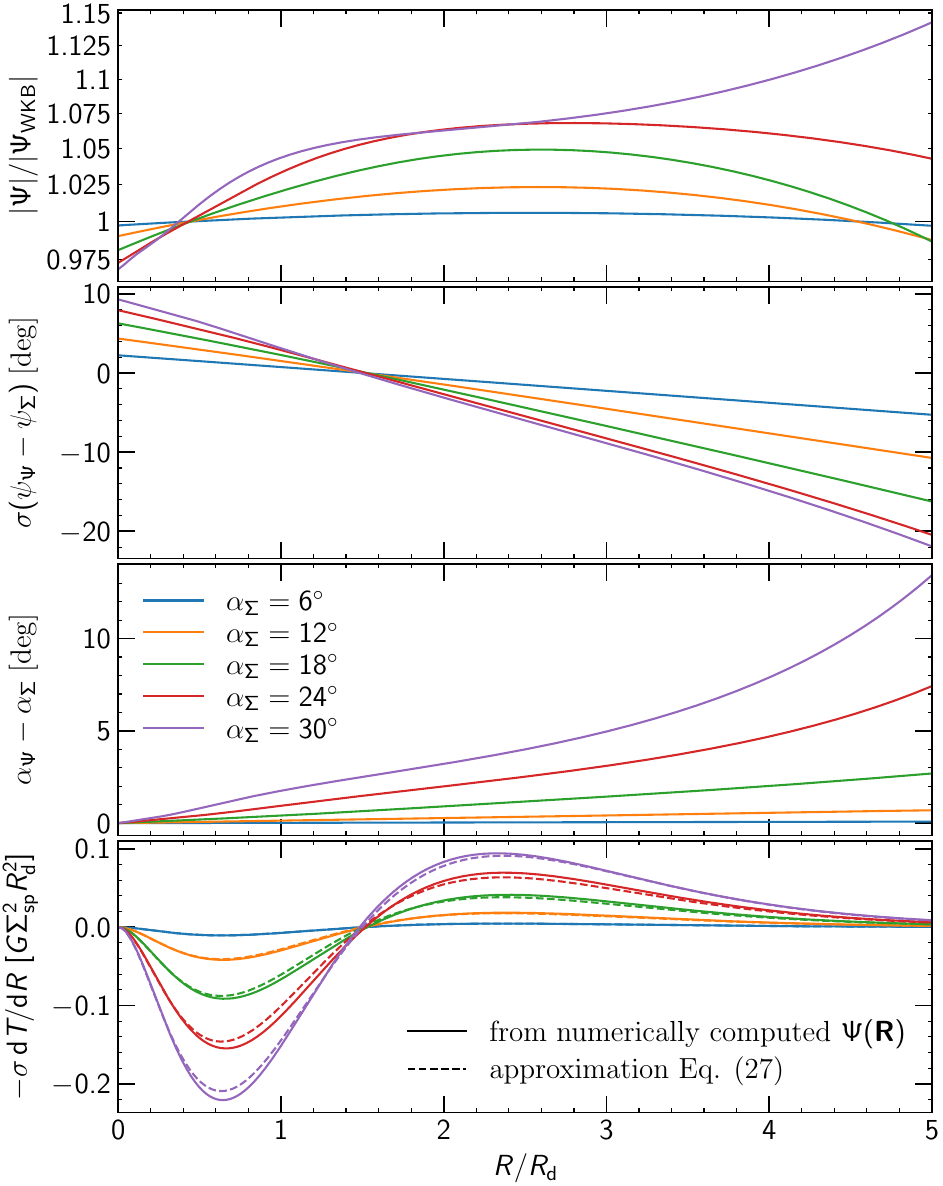}
    \vspace*{-2mm}
    \caption{
    Properties of a spiral with density~\eqref{eq:Sigma:exp:target} and its (numerically computed) potential. Radial profiles of the amplitude ratio to the first-order tight-winding approximation (Eqs.~\ref{eq:WKB} and~\ref{eq:WKB:k:sin}), the offsets of phase and pitch angle between the potential and density, and the torque per annulus ($<0$ if angular momentum is lost for stars in a galaxy with a trailing spiral).}
    \label{fig:exp:Pot:num}
\end{center}
\end{figure}

\subsubsection{The exact gravitational potential}
\label{sec:exp:num}
Finally, I also numerically computed the actual potential generated by the density~\eqref{eq:Sigma:exp:target} and plot in Fig.~\ref{fig:exp:Pot:num} its amplitude relative to that of the traditional first-order tight-winding approximation, as well as the phase offset and the deviation of the pitch angle between the density and potential. While these relations are apparently not easy to predict by tight-winding approximations for large radii and/or pitch angles, they are sufficiently smooth for interpolation, allowing a cost-efficient numerical implementation.

Again, the pitch of the potential is larger than that of the density and increases with radius at large $\alpha_\Sigma$ and $R$ even more so than predicted by our novel second-order approximation.

In the bottom panel Fig.~\ref{fig:exp:Pot:num}, I plot the contribution $-\sigma\dd T/\dd R$ of each annulus to the total torque. Owing to the increasing phase offset $\delta\psi$ between the potential and density changing sign at $R=\sfrac32 R_{\mathrm{d}}$, the torque is negative (for a trailing spiral) at $R<\sfrac32 R_{\mathrm{d}}$ and positive at $R>\sfrac32 R_{\mathrm{d}}$ (such that total angular momentum remains conserved). 

The approximation~\eqref{eq:tau:novel} only slightly underestimates the local torque (dashed in the bottom panel of Fig.~\ref{fig:exp:Pot:num}) and makes a much better impression than the approximations for the potential on which it is based. Moreover, the differences between Eq.~\eqref{eq:tau:novel} and its parent (Eq.~\ref{eq:tau:WKB} with $k$ from Eq.~\ref{eq:WKB:k:SFA}) are very small (not shown).

\subsection{Radially tapered spirals}
\label{sec:taper}
All the spirals considered so far extend radially over the whole disc, but real spirals may be limited to a radial range or, equivalently, an azimuthal range. Theoretical analysis of orbits perturbed by a spiral pattern suggest that the response of the system is in phase with the spiral only in a region around the co-rotation resonance \citep[e.g.][]{Contopoulos1986}, and $N$-body simulations also show spiral patterns to be confined to a region around co-rotation and limited by the Lindblad resonances \citep{Sellwood2019}. Moreover, in barred galaxies, spiral structures are mostly absent from the bar region.

\begin{figure}
	\includegraphics[width=\columnwidth]{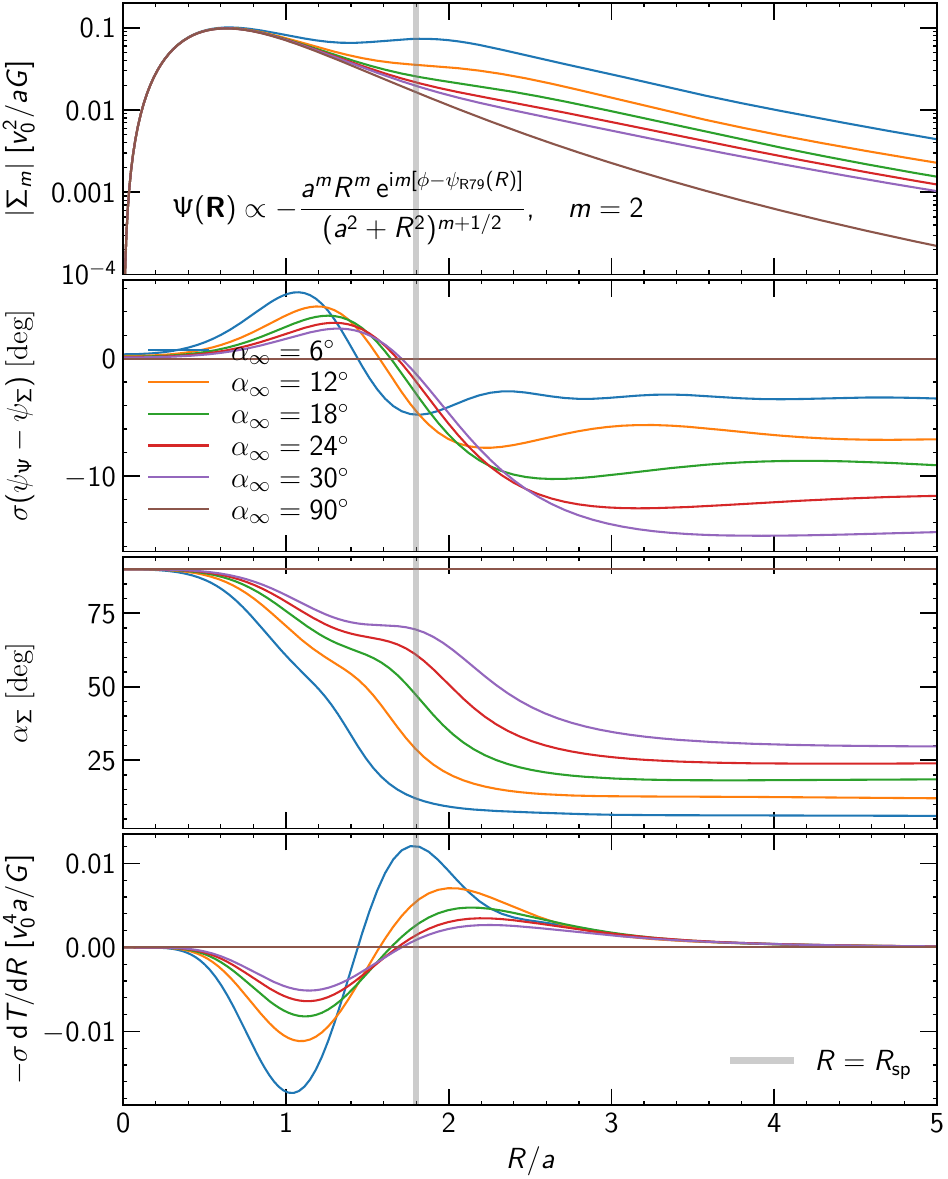}
    \vspace*{-6mm}
    \caption{
    Properties of the (numerically computed) density that generates the potential~\eqref{eq:Psi:Kuzmin} with the phase $\psi=\psi_{\mathrm{R79}}(R)$ (Eq.~\ref{eq:psi:Roberts}) for $R_{\mathrm{sp}}=1.8a$ (grey vertical line) and $N=5$ (as used by \citealt{Roberts1979}) and various values for $\alpha_\infty$ (\citeauthor{Roberts1979} used $\alpha_\infty=20^\circ$). For $\alpha_\infty=90^\circ$, $\psi=0$, and the density is given by Eq.~\eqref{eq:Sigma:Kuzmin}.}
    \label{fig:Roberts}
\end{figure}

\subsubsection{The phase factor of Roberts et al.}
\label{sec:taper:R79}
With the goal to model a spiral that begins just outside a bar, \cite*{Roberts1979} applied the phase 
\begin{align}
    \label{eq:psi:Roberts}
    \psi_{\mathrm{R79}}(R)\equiv \frac\sigma{N\tan\alpha_\infty}\ln\left[1+\left(R/R_{\mathrm{sp}}\right)^N\right]
\end{align}
to the potential
\begin{subequations}
    \label{eqs:Kuzmin:m}
\begin{align}
    \label{eq:Psi:Kuzmin}
    \Psi(\vec{R}) = -v_0^2 \frac{a^{m+1}R^m}{(a^2+R^2)^{m+1/2}}\, \Exp{\imag m(\phi-\psi)},
\end{align}
with $m=2$. At $R\ll R_{\mathrm{sp}}$, this phase function evaluates to zero and the potential perturbation stays aligned ($\alpha=90^\circ$) as for a bar, but at $R>R_{\mathrm{sp}}$, the potential perturbation becomes spiral-like, reaching a pitch angle of $\alpha\to\alpha_\infty$ at $R\gg R_{\mathrm{sp}}$ with the sharpness of the transition determined by the parameter $N$. In case of a constant phase $\psi$,~\eqref{eq:Psi:Kuzmin} is the potential of a razor-thin perturbation with density\footnote{This razor-thin model can be derived from the \cite{Kuzmin1956} disc $\Phi_0$ as $(\partial_x-\imag\partial_y)^m\Phi_0$ and analogously for the density.} 
\begin{align}
    \label{eq:Sigma:Kuzmin}
    \Sigma(\vec{R})=\frac{v_0^2}{2\pi G} \frac{(2m+1)a^{m+2}R^m}{(a^2+R^2)^{m+3/2}}\, \Exp{\imag m(\phi-\psi)},
\end{align}
\end{subequations}
but for the phase~\eqref{eq:psi:Roberts} no formula for the associated density exists (and \citealt{Roberts1979} did not attempt to estimate it). I computed it numerically (for the same parameter values as used by \citeauthor{Roberts1979}) and plot in Fig.~\ref{fig:Roberts} its amplitude, phase, and pitch angle. The latter switches indeed from $\alpha=90^\circ$ at small $R$ to $\alpha=\alpha_\infty$ at large $R$ as intended, albeit perhaps not as cleanly as for the potential: The bar region ($R<R_{\mathrm{sp}}=1.8s$ indicated by grey lines in Fig.~\ref{fig:Roberts}) is significantly pitched, and most of the torque exchange occurs within this region. 

The density declines like $\Sigma\propto\Psi_m/R\propto R^{-4}$, as expected from the tight-winding approximation~\eqref{eq:WKB}, and less steeply than for the bar-like case ($\alpha_\infty=90^\circ$), i.e.\ $\Sigma\propto R^{-5}$ (Eq.~\ref{eq:Sigma:Kuzmin}). The unperturbed ($m=0$) component of the model used by \citeauthor{Roberts1979} was also of type~\eqref{eqs:Kuzmin:m}, such that the relative amplitude of the spiral declined like $R^{-4}/R^{-3}=1/R$.

At least two recent studies applied the phase~\eqref{eq:psi:Roberts} with $N=100$ to the potential~\eqref{eq:Psi:exp:simple}, with the intention of having a spiral pattern only at $R>R_{\mathrm{sp}}$, but unaware that at $R<R_{\mathrm{sp}}$ this generates a bar-like perturbation (in addition to any other explicitly modelled bar) with significant $\sigma\psi'_\Sigma<0$ (twisting the wrong way) near $R=R_{\mathrm{sp}}$. I abstain from plotting these nonsensical models.

\begin{figure}
	\includegraphics[width=\columnwidth]{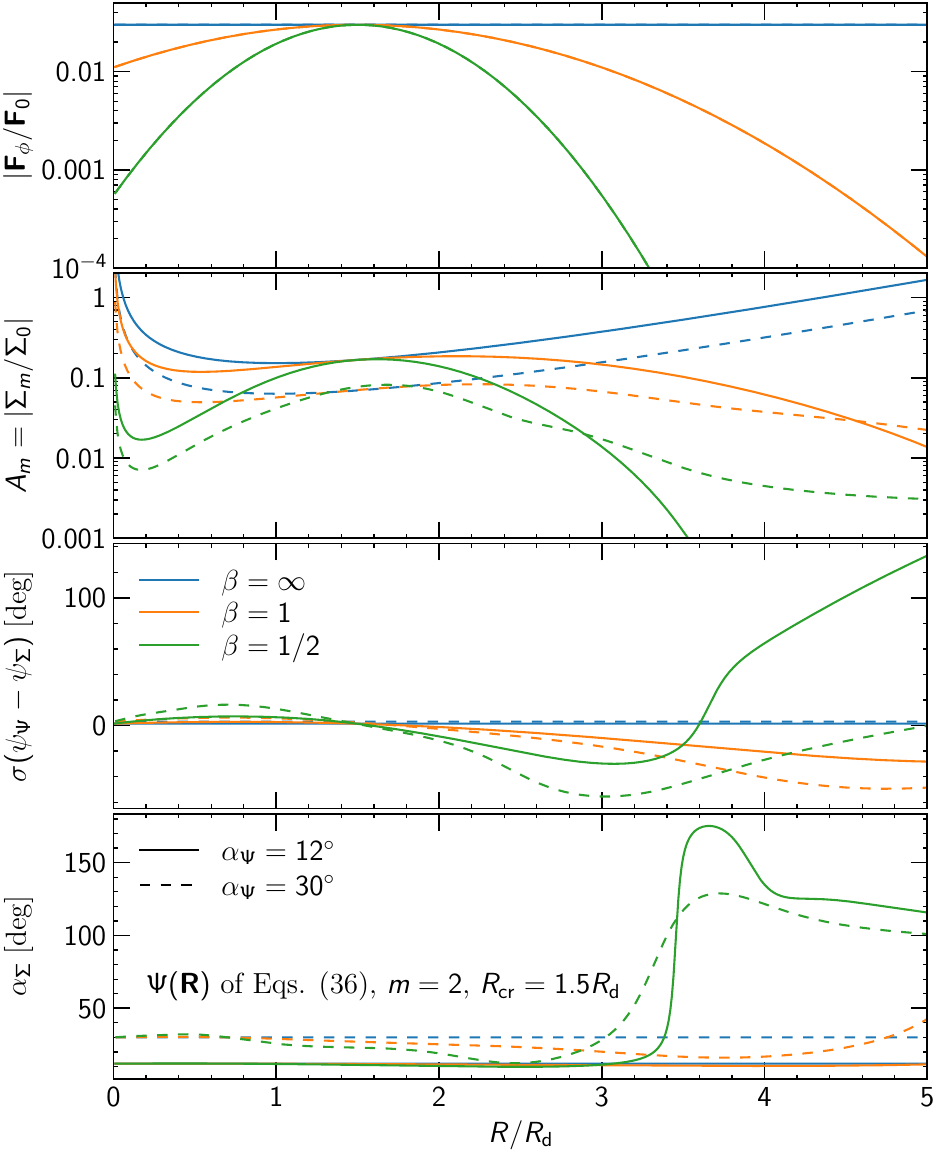}
    \vspace*{-5mm}
    \caption{
    Assessing the model~\eqref{eqs:H24} used by \cite{Hamilton2024} for the same parameters as used by these authors. Plotted as function of radius from top to bottom, I show the relative perturbation strength of the forces, the relative amplitude $A_{m=2}$ of the density perturbation (assuming the unperturbed disc is maximal with scale radius related to the co-rotation radius of the spiral as $R_{\mathrm{cr}}=1.5R_{\mathrm{d}}$), the phase offset between the potential and density, and the density pitch angle.}
    \label{fig:Hamilton:beta}
\end{figure}

\subsubsection{The spirals of Hamilton et al.}
\label{sec:taper:H24}
\cite*{Hamilton2024} used a model with the potential being a logarithmic spiral,
\begin{subequations}  
    \label{eqs:H24}
\begin{align}
\label{eq:H24:Psi}
    \Psi(\vec{R}) &= \frac\eta m v_0^2 \,B_{\mathrm{H24}}(R,\beta) \, \Exp{\imag m[\phi-\lambda\ln(R/R_0)]}
\end{align}
with
\begin{align}
\label{eq:H24:B}
    B_{\mathrm{H24}}(R,\beta) &\equiv \exp\bigg(\!-\frac{(R-R_{\mathrm{cr}})^2}{2R_\beta^2}\bigg),
\end{align}
\end{subequations}
where $R_\beta=\beta\sqrt{2}R_{\mathrm{cr}}/m$ with $R_{\mathrm{cr}}$ the co-rotation radius of the spiral pattern. \citeauthor{Hamilton2024} modelled the $m=0$ component of the total potential as $\Psi=v_0^2\ln R$, such that $\eta$ is the (maximum of the) ratio of the tangential force to the unperturbed force. Moreover, for this model the epicycle and circular frequencies are related as $\kappa=\sqrt{2}\Omega$, such that the Lindblad resonances are located at $R=(1\pm\sqrt{2}/m)R_{\mathrm{cr}}$, and for $\beta=1$ the standard deviation of the potential envelope $B_{\mathrm{H24}}$ equals the distance to the Lindblad resonances. \citeauthor{Hamilton2024} considered $\beta=\sfrac12$, $\beta=1$, and $\beta=\infty$. The latter case is identical to the scale-invariant spirals~\eqref{eq:Psi:From:Surf:scale-invariant} with $\gamma=1$, while for finite $\beta$, I computed $\Sigma(\vec{R})$ numerically.

A common measure of the spiral strength is the ratio $A_m=|\Sigma_m/\Sigma_0|$ of the density amplitudes of the perturbation and the unperturbed ($m=0$) disc. I obtained (a lower limit for) $A_m$ by assuming a maximum disc for the $m=0$ component, i.e.\ Eq.~\eqref{eq:disc:exp} with $\Sigma_{\mathrm{c}} = 0.411\, v_0^2/GR_{\mathrm{d}}$, such that the maximum circular speed generated by the disc alone (without the dark-matter halo) equals $v_0$.

In Fig.~\ref{fig:Hamilton:beta}, the resulting $A_{m=2}$ is plotted (in the second panel from top) for the parameter choices used by \citeauthor{Hamilton2024} and $R_{\mathrm{cr}}=1.5R_{\mathrm{d}}$, while the lower panels show the phase offset and density pitch angle as in previous figures. For $\beta=\infty$, which is their favourite model, $A_{m=2}$ has an inverted profile with a minimum at $R=R_{\mathrm{d}}$. Towards smaller and larger radii, it increases and eventually exceeds unity, implying negative density. The finite-$\beta$ models do not suffer from this issue (except at $R\ll R_{\mathrm{d}}$), but their spiral strength $A_{m=2}$ is not symmetric with respect to $R=R_{\mathrm{cr}}$ (unlike the potential amplitude) and peaks at $R>R_{\mathrm{cr}}$.

For the model $\beta=\sfrac12$ with $\alpha=30^\circ$, the density pitch angle decreases at $R>R_{\mathrm{cr}}$, implying that the spiral becomes tighter than anticipated by \citeauthor{Hamilton2024}. The implications of all these properties for the results of the study by \cite{Hamilton2024} are discussed in Section~\ref{sec:discuss:poor}.

\begin{figure*}
	\includegraphics[width=\columnwidth]{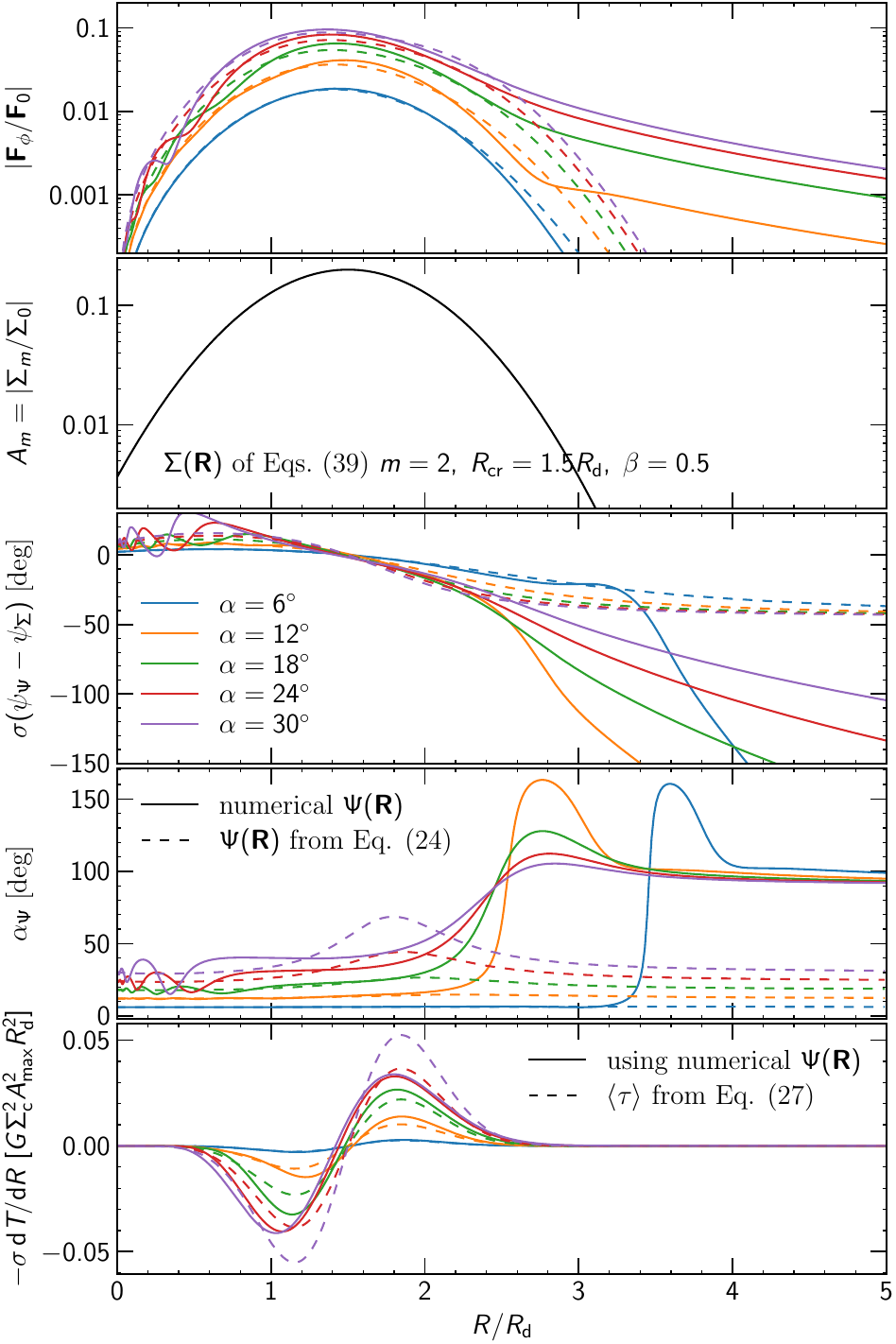}\hfill
	\includegraphics[width=\columnwidth]{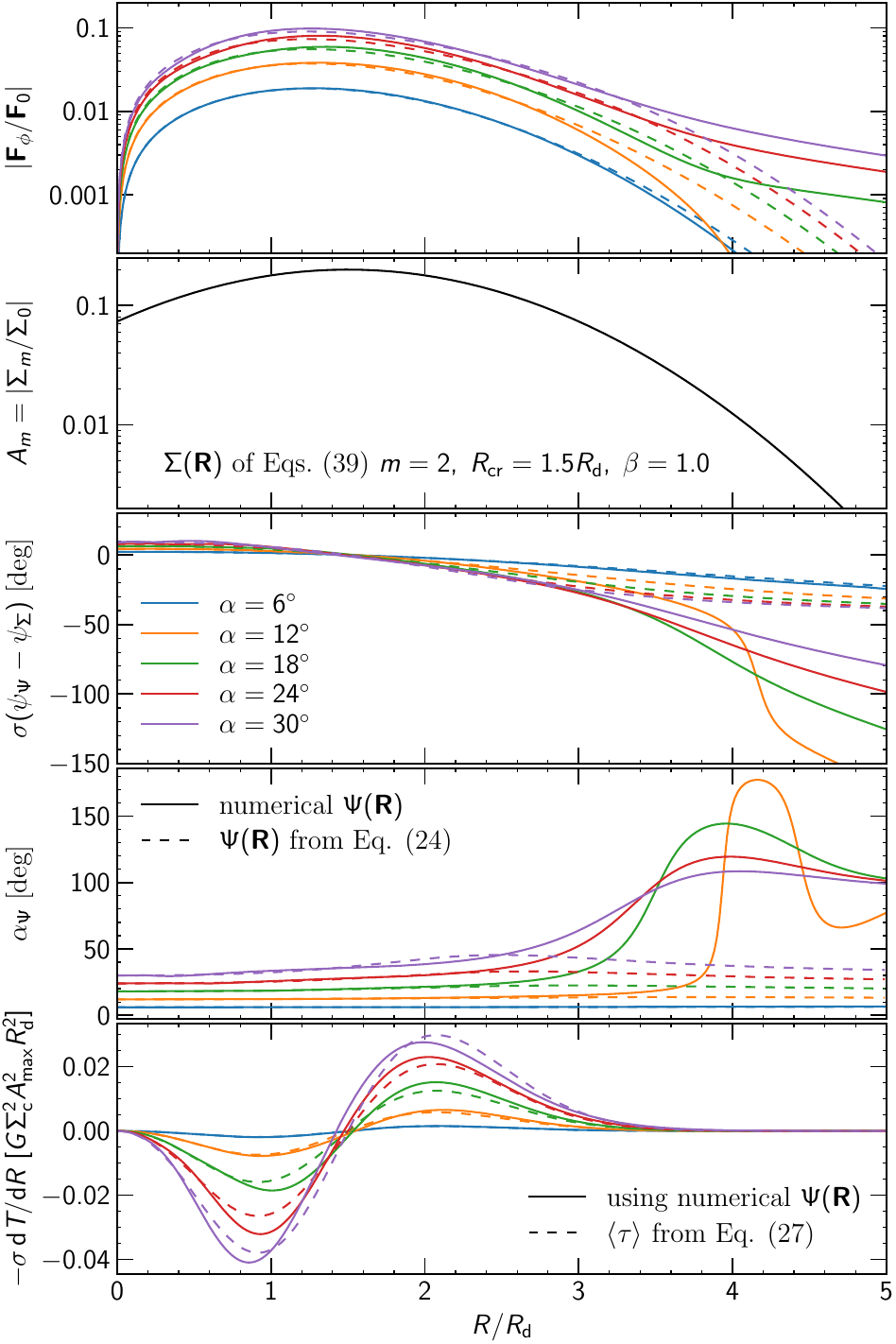}
    \vspace*{-2mm}
    \caption{
    Similar to Fig.~\ref{fig:Hamilton:beta}, but for the density models of Eq.~\eqref{eq:Sigma:exp:taper} with constant pitch angle $\alpha$ and relative perturbation density-amplitude $A_m(R)=A_{\max}B_{\mathrm{H24}}(R,\beta)$, i.e.,\ the same profile as the relative perturbation force-amplitude for the model of \citeauthor{Hamilton2024} (\citeyear{Hamilton2024}, shown in Fig.~\ref{fig:Hamilton:beta}) for $\beta=\sfrac12$ (\textbf{left}) and $\beta=1$ (\textbf{right}). Differently from that figure, the pitch angle of the potential is plotted. The estimates for the potential from the second-order tight-winding approximation~\eqref{eq:WKB:novel} are also shown, as is the torque per annulus (\textbf{bottom}) and its approximation~\eqref{eq:tau:novel}.}
    \label{fig:expDisc:beta}
\end{figure*}

\subsubsection{Spirals with simple density models}
\label{sec:taper:real}
All the models considered so far (with the exception of Fig.~\ref{fig:exp:Pot:num}) were specified via the spiral potential, which is required for dynamical modelling. Unfortunately, as we have seen, simple potentials tend to correspond to spiral density patterns that deviate from the target, in particular, if the spiral pattern is restricted in radius. Instead, I now consider simple models in the density and numerically compute the corresponding potential.

I still assumed that the unperturbed ($m=0$) density is of the form~\eqref{eq:disc:exp} and that the relative amplitude of the spiral perturbation is $A_m(R)=|\Sigma_m|/\Sigma_0$. The spiral perturbation then has the density
\begin{subequations}
\begin{align}
    \label{eq:Sigma:exp:taper}
    \Sigma(\vec{R}) = \Sigma_{\mathrm{c}}\, A_m(R)\, \Exp{-R/R_{\mathrm{d}}+\imag m[\phi-\lambda\ln(R/R_0)]}.
\end{align}
 For constant $A_m$, this is identical to the target density~\eqref{eq:Sigma:exp:target} from the previous section, whose potential was presented in Fig.~\ref{fig:exp:Pot:num}. Instead, I now consider simple non-constant amplitude functions of the form 
\begin{align}
    A_m(R) = A_{\max}\,B_{\mathrm{H24}}(R,\beta)
\end{align}   
\end{subequations}
and plot in Fig.~\ref{fig:expDisc:beta} for $\beta=\sfrac12$ and 1 ($\beta=\infty$ gives the model shown in Fig.~\ref{fig:exp:Pot:num}) the same quantities as for the models of \citeauthor{Hamilton2024} in Fig.~\ref{fig:Hamilton:beta}, plus the torque per annulus.

A comparison of these two figures shows many similarities: (i) The relative force perturbation peaks at smaller radii than the relative density perturbation, although this effect is weaker for our more realistic density models, where it is hardly present for $\beta=\sfrac12$. (ii) At radii where $A_m$ is significant (not much smaller than its maximum), the phase offset $\delta\psi\equiv\psip-\psis$ decreases slowly with radius (already seen in Fig.~\ref{fig:exp:Pot:num}), and the potential pattern becomes more open with an outward-increasing pitch angle (although this effect is much weaker for the density models). Finally, (iii) at large radii where  $A_m$ is insignificant, the potential decays like $R^{-1-m}$ with $\alpha_\Psi=90^\circ$. This is simply the expected behaviour for the outer field of the inner spiral \citep[e.g.][Eq.~2.93]{BT2008} without any significant local perturbation, and it is therefore also impossible to predict via the tight-winding approximations (dashed in Fig.~\ref{fig:expDisc:beta}). In the transition regions, at the edges of the spiral, the potential perturbation even twists backwards, as indicated by the pitch angle of the potential exceeding $90^\circ$.

In the bottom panels of Fig.~\ref{fig:expDisc:beta}, I also plot the net torque per annulus. The general pattern, already seen in Fig.~\ref{fig:exp:Pot:num} for a spiral without radial truncation, of outward angular-momentum transport by a trailing spiral is also present for these radially tapered spirals. The approximation~\eqref{eq:tau:novel} (dashed) is not perfect, but gives a reasonable estimate for this torque, in particular, for spirals with a larger radial extent and/or smaller pitch angles. Interestingly, the total torque at $-\sigma\av\tau<0$, i.e.\ the total amount of angular momentum that is transported non-locally per unit time, is comparable for the models of $\beta=\sfrac12$ and $\beta=1$ (and the same $\alpha$): For the shorter spiral pattern, the torque is limited to a smaller region, which is compensated for by a higher torque density.

\begin{figure*}
    \hfill\includegraphics[width=17cm]{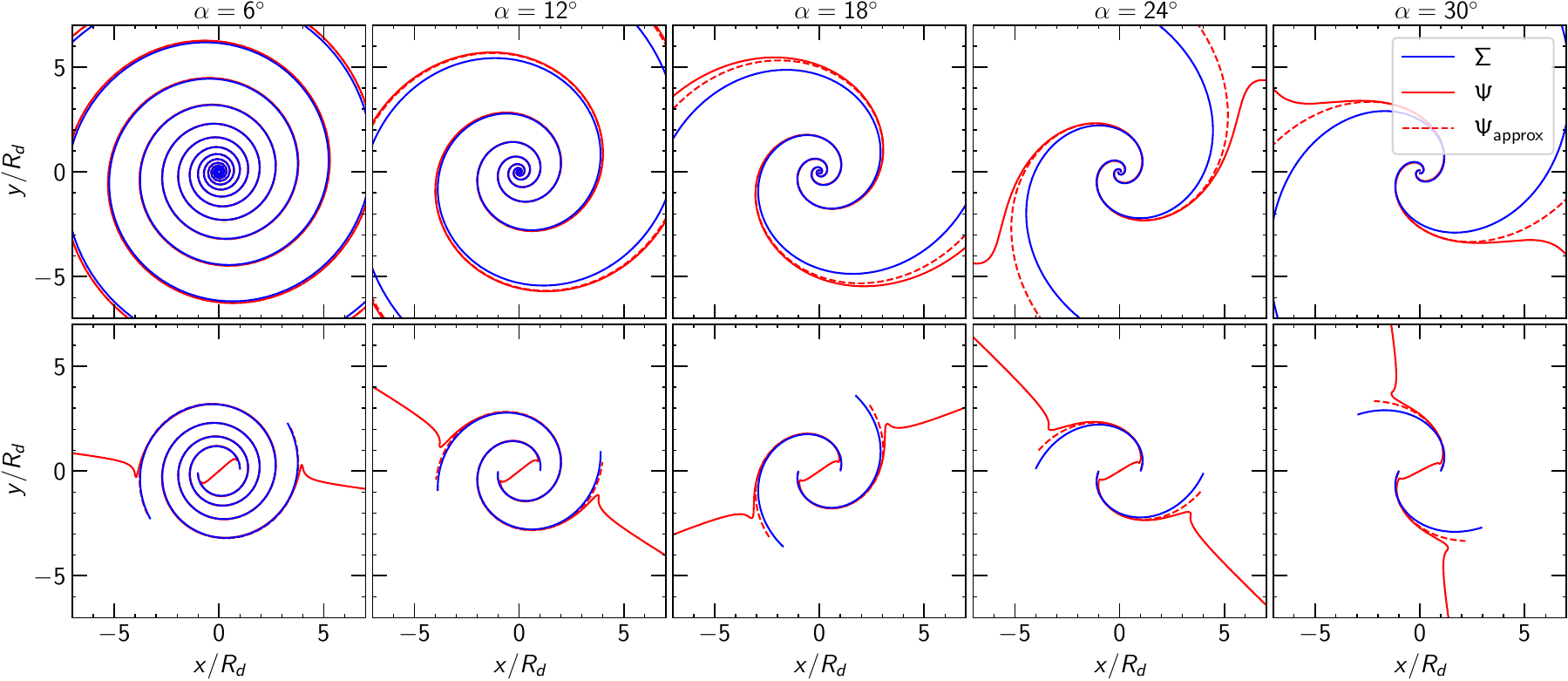}\hfill
    \vspace*{-2mm}
    \caption{
    Crest lines of the density $\Sigma$ (blue) and potential $\Psi$ (red, partly concealed by those of the density) for logarithmic $m=2$ spirals with a pitch angle $\alpha$ as indicated (top) and an exponentially declining radial density-amplitude (i.e.\ constant relative amplitude $A_m=|\Sigma_m|/\Sigma_0$), either radially unlimited (\textbf{top}) or with sharp inner and outer edges (\textbf{bottom}). I also plot (dashed) crest lines of the potential obtained via the second-order approximation~\eqref{eq:WKB:novel} (to first order, the crest lines of $\Sigma$ and $\Psi$ are identical).}
    \label{fig:crest}
\end{figure*}

\section{Discussion}
\label{sec:discuss}
\subsection{How good is the tight-winding (WKB) approximation?}
\label{sec:discuss:WKB}
Many of the early results on disc stability rely on the first-order tight-winding (or an equivalent) approximation \citep[e.g.][]{Toomre1964, Goldreich1965, JulianToomre1966, Toomre1981}, and various later studies have employed models for the potential of spirals based on this approximation \citep[e.g.][]{Contopoulos1986, Grosbol1993, Grosbol2018, Eilers2020}. Little is known about its accuracy, however. \cite{BT2008} claimed that ``in many situations [it] works fairly well even for $|kR|$ as small as unity'' (i.e.\ pitch angles as large as $60^\circ$ for $m=2$), but failed to give evidence of this claim. When prompted, Scott Tremaine (private communication) pointed to the fact that it works very well for \citeauthor{Kalnajs1971}' (\citeyear{Kalnajs1971}) spiral models, as I have shown in Section~\ref{sec:WKB:cmp}. These models are exceptional, however, in that the density and potential have the same phase, which the (first-order) tight-winding approximation always predicts, but which does not hold in general, nor for realistic spirals.

I assessed the tight-winding approximation for the important case of a spiral perturbation with an amplitude that declines exponentially, like the unperturbed surface density itself, such that the relative strength of the spiral is constant with radius. For this case, the amplitude of the perturbing potential is fairly well approximated, in particular, if the non-standard form~\eqref{eq:WKB:k:sin} for $k$ is used. However, the first-order approximation does not give a phase offset $\delta\psi$ between the potential and density, and hence, it fails to predict the radially increasing pitch of the potential. If the spiral perturbation has a finite radial extent, the tight-winding approximation only works within the spiral region and fails at its edges and outside. 

\subsection{An improved tight-winding approximation}
\label{sec:discuss:WKB:improved}
\cite{Kalnajs1971} provided the exact potential of logarithmic spirals whose density amplitude declines like $S_{\!m}\propto R^{-3/2}$. I generalised these models to $S_{\!m}\propto R^{-\gamma}$, where the exponent $\gamma$ is a free parameter. These appear to be the only flat spiral models whose potential is known in closed form. For $\gamma\neq\sfrac32$, the phase of the potential in the equatorial plane, $\Psi(\vec{R})$, differs from that of the density by a constant offset $\delta\psi=\psip-\psis$ (such that the pitch angle is still the same), which is approximately proportional to $\gamma-\sfrac32$.

By locally approximating an arbitrary spiral pattern by such a scale-invariant model, I obtained the second-order tight-winding approximation~\eqref{eq:WKB:novel}
\begin{align}
    \nonumber
    \Psi(\vec{R}) &\approx -2\pi G \frac{\sin\alpha}{m}\left[1+
    \imag\sigma\frac{\sin2\alpha}{2m}\left(\gamma-\tfrac32\right)\right]R\,\Sigma(\vec{R}),
\end{align}
where $\alpha=\alpha(R)$ is the local pitch angle of the density and $\gamma=\gamma(R)=-\dd\ln S_{\!m}/\dd \ln R$ the power-law slope of its amplitude, while $\sigma=\sign(\dd\psis/\dd R)$ gives the sense of the spiral twist (for a trailing spiral $\sigma=-\sign(\bar{v}_\phi)$). This novel tight-winding approximation is much better than the traditional first-order approach. In particular, it predicts the phase offset $\delta\psi$ and pitch of the potential with reasonable accuracy.

\subsection{The potential of typical spiral perturbations}
\label{sec:discuss:pot}
As the surface density of spiral galaxies typically declines nearly exponentially with radius, so does the amplitude of its spiral perturbation (except at its inner and outer edges). In particular, the amplitude declines faster than a power law, and as a consequence, the phase offset $\delta\psi$ is no longer constant. We may estimate it from the second-order approximation~\eqref{eq:WKB:novel} as
\begin{align}
    \label{eq:delta:psi:approx:1}
    \delta\psi(R) 
    &\approx -\frac\sigma m\arctan\left[\frac{\sin2\alpha}{2m}\left(\gamma-\tfrac32\right)\right] 
    \simeq -\frac{\sigma}{m^2} \alpha \left(\gamma-\tfrac32\right).
\end{align}
Typically, the density amplitude is shallower than $S_{\!m}\propto R^{-3/2}$ at small radii and steeper at larger radii, such that the potential trails the density at small $R$ and leads it at large $R$ (for a trailing spiral). For a purely exponential profile with scale radius $R_{\mathrm{d}}$, the transition occurs at $R=\sfrac32 R_{\mathrm{d}}$. 

I demonstrate this in Fig.~\ref{fig:crest}, which plots the crest lines (the positions of the density maxima and potential minima at each radius) for logarithmic spirals with an exponentially declining amplitude and various pitch angles (for the density). The solid red curves show the exact potential, and the dashed curves were obtained by the approximation~\eqref{eq:WKB:novel}, which out to $R=5R_{\mathrm{d}}$ predicts the actual potential very well. Beyond this radius, this local approximation fails because the potential depends ever more on the inner rather than the local spiral.

The bottom panels of Fig.~\ref{fig:crest} plot the same, but for a spiral that has a non-zero (and exponentially declining) amplitude only at $R_{\mathrm{d}} < R < 4R_{\mathrm{d}}$. In most of this range, the potential closely follows the density, as predicted by the tight-winding approximations, but towards the edges and outside of the spiral range, the potential aligns itself to a constant phase and decays like $\Psi\propto R^m$ at $R\to0$ and $\Psi\propto R^{-m-1}$ at $R\to\infty$, as expected for the outer potential of an $m$-fold  perturbation.

Galactic spirals tend to have an outward-decreasing pitch angle \citep{Savchenko2013}, which implies via Eq.~\eqref{eq:delta:psi:approx:1} that the phase offset does not increase as much as for $\alpha=\,$const.

\subsection{Net torque and non-local angular momentum transport}
The potential of a spiral perturbation exerts a torque $-\partial\Phi/\partial\phi$ per unit mass. For most stars, this torque averages out, but for stars in orbital resonance, the torque accumulates and causes significant angular-momentum changes $\delta J_\phi$. These changes cause radial migration \citep{Sellwood2002} by (i) advective angular-momentum transport caused by local diffusion of $J_\phi$, as characterised by $\langle(\delta J_\phi)^2\rangle$, and (ii) non-local angular-momentum transport described directly by $\langle\delta J_\phi\rangle$.

The sign of $\delta J_\phi$ depends on the relative orbital phase of the star to the spiral perturbation, which is not completely random, as the spiral pattern is made up of the very same stars. As a result, there remains a net torque at each radius, which is second order in both the spiral amplitude and pitch angle, and the sign of which depends on whether the potential of the perturbation trails ($\sigma\delta\psi>0$) or leads ($\sigma\delta\psi<0$) the density. For trailing spirals, $\sigma=-\sign(\bar{v}_\phi)$ and the net torque reduces stellar angular momenta at small radii and increases them at large radii, constituting a non-local outward transport of angular momentum by gravitational torques \citep{Zhang1996}.

Thus, every trailing spiral perturbation transports angular momentum outwards, not only via local diffusion, but also by a net gravitational torque at each radius. Remarkably, this conclusion can be derived analytically (in Sections~\ref{sec:torque:approx:WKB} and~\ref{sec:torque:approx:pot}) without considering orbital dynamics. It implies that spirals alter the state of the disc irreversibly.

\subsection{Implications of poor spiral models}
\label{sec:discuss:poor}
The results of models using inadequate gravitational potentials for the spiral pattern are obviously compromised. Unfortunately, these shortcomings are not as uncommon as might be hoped. I give two recent examples.

With the intention to model Milky Way spirals, \cite{Eilers2020} used a gravitational potential that was meant to be the first-order tight-winding approximation~\eqref{eq:Psi:exp:simple}, but deviated by a factor $R\tan\alpha/mh_z$ with $h_z=1\,$kpc. This, however, gives the tight-winding approximation for a density amplitude that differs from the intended exponential model by the same factor, i.e.\ the relative amplitude $A_m=|\Sigma_m|/\Sigma_0$ of the spiral perturbation increases linearly with radius. This is certainly unphysical and implies that the associated dynamical model of the non-axisymmetric stellar kinematics is inadequate (\citeauthor{Eilers2020} also neglected the gravity from the Milky Way bar, which renders their model even less adequate).

As already shown in Section~\ref{sec:taper:H24}, the models employed by \cite{Hamilton2024} also used unrealistic profiles of the density amplitude of the spiral perturbation, in particular, their model with $\beta=\infty$ in Eqs.~\eqref{eqs:H24}, which \citeauthor{Hamilton2024}  favoured because it would resemble the model of \cite{Eilers2020}. Their models $\beta=\sfrac12$ and $\beta=1$ have more reasonable density profiles and induce much less radial heating, which agrees well with the situation that is observed for the Milky Way. The claim of \citeauthor{Hamilton2024} that the Milky Way stellar disc is dynamically colder than expected is therefore unfounded.

\subsection{The vertical dimension}
\label{sec:discuss:z}
In this study, only flat (razor-thin) spiral perturbations have been considered, and of these, only their gravitational potential in the equatorial plane. This is often sufficient for modelling horizontal disc dynamics, at least to first order. 

While the discs and their spiral perturbation of real galaxies are thin in the sense that the scale height $h_{\mathrm{z}}$ is much smaller than the scale length $R_{\mathrm{d}}$, $h_{\mathrm{z}}$ can become significant compared to the radial wavelength of a spiral perturbation. In the first-order tight-winding approximation, the potential decays as $\Exp{-k|z|}$ away from $z=0$, where $k=m/R\sin\alpha$, and convolving this with the vertical profile $\propto\Exp{-|z|/h_{\mathrm{z}}}$ reduces the central potential by the factor $(1+h_{\mathrm{z}}k)^{-1}$. This reduction is larger at smaller $R$ and smaller $\alpha$ (more tightly wound spirals), and implies that razor-thin models overestimate the gravity of spiral perturbations by $\sim(1+h_{\mathrm{z}}k)$.

To avoid this and model the vertical extent of galactic discs and their spiral perturbations more accurately, full three-dimensional models for these perturbations and their potentials are required. These are the subject of an ongoing study.

\section{Conclusions}
\label{sec:conclude}
I studied the gravitational potential of flat (razor-thin) spiral perturbations by means of an accurate numerical solution and compared them to the widely used tight-winding (or WKB) approximation. I also extended \citeauthor{Kalnajs1971}' (\citeyear{Kalnajs1971}) scale-invariant spirals to general power-law amplitudes $|\Sigma_m|\propto R^{-\gamma}$ (\citeauthor{Kalnajs1971}'s spirals have $\gamma=\sfrac32$). I studied various simple models for spiral potentials or densities, most of which were logarithmic spirals (constant pitch angle $\alpha$). My main results are as follows.
\begin{itemize}
    \item For typical $m=2$ spirals with exponentially declining amplitude (and hence, with constant relative amplitude), the traditional (first-order) tight-winding approximation gives $\lesssim10\%$ errors for pitch angles $\alpha\lesssim20^\circ$, and deteriorates beyond that.
    \item Models for spiral potentials used in the literature are typically based on the first-order tight-winding approximation and inherit all its weaknesses, implying that the actually implied density deviates from the intended target, sometimes significantly so.
    \item A second-order tight-winding approximation (Eq.~\ref{eq:WKB:novel}), based on locally approximating the spiral perturbation by a scale-invariant spiral (rather than a plane wave), is significantly better and can predict the phase offset $\delta\psi$ of the potential from the density (unlike the first-order approximations, which always give $\delta\psi=0$).
    \item Towards the inner or outer edge of a spiral perturbations, all tight-winding approximations fail. Beyond the edges, the gravitational potential of the spiral resembles that of a weak bar. Hence, to model these radially limited spiral perturbations, approximate techniques cannot be used, but numerical treatment is required.
    \item Generally, the gravitational potential for trailing spirals trails the density at small radii and leads it at large radii, with the transition at the radius at which the density amplitude declines like $R^{-3/2}$. These phase offsets result in gravitational torques that transport angular momentum outwards and change the state of the disc irreversibly. Eq.~\eqref{eq:tau:novel} gives a simple but reasonably accurate approximation for the average torque density at each radius.
    \item I developed an efficient numerical method for computing the gravitational potential of spiral perturbations for arbitrary multiplicity $m$, phase functions $\psi(R)$, and amplitudes $S_{\!m}(R)$. Implementations in \texttt{C++} and \texttt{python} are made available.
\end{itemize}

\begin{acknowledgements}
I thank Howard Cohl for discussing the Laplace coefficients and for sharing Kalnajs' letter to him, Scott Tremaine for discussing the computation of the Laplace coefficients and the tight-winding approximation, Agris Kalnajs for opening my eyes to gravitational torques by pointing out the related problem of the scale-invariant spirals with $\gamma\neq\sfrac32$, and the anonymous reviewer for a thorough reading and a prompt, encouraging, and helpful report.
\end{acknowledgements}


\bibliographystyle{mnras}
\bibliography{aa55980-25}

\begin{thebibliography}{}
\makeatletter
\relax
\def\mn@urlcharsother{\let\do\@makeother \do\$\do\&\do\#\do\^\do\_\do\%\do\~}
\def\mn@doi{\begingroup\mn@urlcharsother \@ifnextchar [ {\mn@doi@} {\mn@doi@[]}}
\def\mn@doi@[#1]#2{\def\@tempa{#1}\ifx\@tempa\@empty \href {http://dx.doi.org/#2} {doi:#2}\else \href {http://dx.doi.org/#2} {#1}\fi \endgroup}
\def\mn@eprint#1#2{\mn@eprint@#1:#2::\@nil}
\def\mn@eprint@arXiv#1{\href {http://arxiv.org/abs/#1} {{\tt arXiv:#1}}}
\def\mn@eprint@dblp#1{\href {http://dblp.uni-trier.de/rec/bibtex/#1.xml} {dblp:#1}}
\def\mn@eprint@#1:#2:#3:#4\@nil{\def\@tempa {#1}\def\@tempb {#2}\def\@tempc {#3}\ifx \@tempc \@empty \let \@tempc \@tempb \let \@tempb \@tempa \fi \ifx \@tempb \@empty \def\@tempb {arXiv}\fi \@ifundefined {mn@eprint@\@tempb}{\@tempb:\@tempc}{\expandafter \expandafter \csname mn@eprint@\@tempb\endcsname \expandafter{\@tempc}}}

\bibitem[\protect\citeauthoryear{{Abramowitz} \& {Stegun}}{{Abramowitz} \& {Stegun}}{1972}]{AS}
{Abramowitz} M.,  {Stegun} I.~A.,  1972, {Handbook of Mathematical Functions}.
Dover Publications, New York

\bibitem[\protect\citeauthoryear{{Binney} \& {Tremaine}}{{Binney} \& {Tremaine}}{2008}]{BT2008}
{Binney} J.,  {Tremaine} S.,  2008, {Galactic Dynamics}, 2. edn.
Princeton University Press, Princeton NJ

\bibitem[\protect\citeauthoryear{{Chugunov}, {Marchuk}  \& {Savchenko}}{{Chugunov} et~al.}{2025}]{Chugunov2025}
{Chugunov} I.~V.,  {Marchuk} A.~A.,   {Savchenko} S.~S.,  2025, \mn@doi [Galaxies] {10.3390/galaxies13020044}, \href {https://ui.adsabs.harvard.edu/abs/2025Galax..13...44C} {13, 44}

\bibitem[\protect\citeauthoryear{{Contopoulos} \& {Grosb{\o}l}}{{Contopoulos} \& {Grosb{\o}l}}{1986}]{Contopoulos1986}
{Contopoulos} G.,  {Grosb{\o}l} P.,  1986, \aap, \href {https://ui.adsabs.harvard.edu/abs/1986A&A...155...11C} {155, 11}

\bibitem[\protect\citeauthoryear{{Eilers}, {Hogg}, {Rix}, {Frankel}, {Hunt}, {Fouvry}  \& {Buck}}{{Eilers} et~al.}{2020}]{Eilers2020}
{Eilers} A.-C.,  {Hogg} D.~W.,  {Rix} H.-W.,  {Frankel} N.,  {Hunt} J. A.~S.,  {Fouvry} J.-B.,   {Buck} T.,  2020, \mn@doi [\apj] {10.3847/1538-4357/abac0b}, \href {https://ui.adsabs.harvard.edu/abs/2020ApJ...900..186E} {900, 186}

\bibitem[\protect\citeauthoryear{{Goldreich} \& {Lynden-Bell}}{{Goldreich} \& {Lynden-Bell}}{1965}]{Goldreich1965}
{Goldreich} P.,  {Lynden-Bell} D.,  1965, \mn@doi [\mnras] {10.1093/mnras/130.2.125}, \href {https://ui.adsabs.harvard.edu/abs/1965MNRAS.130..125G} {130, 125}

\bibitem[\protect\citeauthoryear{{Gough}}{{Gough}}{2011}]{GSL}
{Gough} B.,  ed. 2011, GNU scientific library : reference manual, 3. edn.
Network Theory, Bristol

\bibitem[\protect\citeauthoryear{{Grosb{\o}l}}{{Grosb{\o}l}}{1993}]{Grosbol1993}
{Grosb{\o}l} P.,  1993, \mn@doi [\pasp] {10.1086/133212}, \href {https://ui.adsabs.harvard.edu/abs/1993PASP..105..651G} {105, 651}

\bibitem[\protect\citeauthoryear{{Grosb\o{}l} \& {Carraro}}{{Grosb\o{}l} \& {Carraro}}{2018}]{Grosbol2018}
{Grosb\o{}l} P.,  {Carraro} G.,  2018, \mn@doi [A\&A] {10.1051/0004-6361/201833755}, \href {https://ui.adsabs.harvard.edu/abs/2018A%26A...619A..50G} {619, A50}

\bibitem[\protect\citeauthoryear{{Hagihara}}{{Hagihara}}{1972}]{CeMe1972}
{Hagihara} Y.,  1972, {Celestial mechanics. Vol.2: Perturbation theory}.
MIT Press, Cambridge MS

\bibitem[\protect\citeauthoryear{{Hamilton}, {Modak}  \& {Tremaine}}{{Hamilton} et~al.}{2024}]{Hamilton2024}
{Hamilton} C.,  {Modak} S.,   {Tremaine} S.,  2024, \mn@doi [arXiv e-prints] {10.48550/arXiv.2411.08944}, \href {https://ui.adsabs.harvard.edu/abs/2024arXiv241108944H} {p. arXiv:2411.08944}

\bibitem[\protect\citeauthoryear{{Hur{\'e}}}{{Hur{\'e}}}{2005}]{Hure2005}
{Hur{\'e}} J.~M.,  2005, \mn@doi [\aap] {10.1051/0004-6361:20034194}, \href {https://ui.adsabs.harvard.edu/abs/2005A&A...434....1H} {434, 1}

\bibitem[\protect\citeauthoryear{{Julian} \& {Toomre}}{{Julian} \& {Toomre}}{1966}]{JulianToomre1966}
{Julian} W.~H.,  {Toomre} A.,  1966, \mn@doi [\apj] {10.1086/148957}, \href {https://ui.adsabs.harvard.edu/abs/1966ApJ...146..810J} {146, 810}

\bibitem[\protect\citeauthoryear{{Kalnajs}}{{Kalnajs}}{1971}]{Kalnajs1971}
{Kalnajs} A.~J.,  1971, \mn@doi [\apj] {10.1086/150957}, \href {https://ui.adsabs.harvard.edu/abs/1971ApJ...166..275K} {166, 275}

\bibitem[\protect\citeauthoryear{{Kennicutt}}{{Kennicutt}}{1981}]{Kennicutt1981}
{Kennicutt} Jr. R.~C.,  1981, \mn@doi [\aj] {10.1086/113064}, \href {https://ui.adsabs.harvard.edu/abs/1981AJ.....86.1847K} {86, 1847}

\bibitem[\protect\citeauthoryear{{Kuzmin}}{{Kuzmin}}{1956}]{Kuzmin1956}
{Kuzmin} G.~G.,  1956, Astron. Zh., 33, 27

\bibitem[\protect\citeauthoryear{{Lin} \& {Shu}}{{Lin} \& {Shu}}{1964}]{LinShu1964}
{Lin} C.~C.,  {Shu} F.~H.,  1964, \mn@doi [\apj] {10.1086/147955}, \href {https://ui.adsabs.harvard.edu/abs/1964ApJ...140..646L} {140, 646}

\bibitem[\protect\citeauthoryear{{NIST}}{{NIST}}{2024}]{NIST-DLMF}
{NIST} 2024, Digital Library of Mathematical Functions, \url{https://dlmf.nist.gov/}, Release 1.2.3 of 2024-12-15

\bibitem[\protect\citeauthoryear{{Roberts}, {Huntley}  \& {van Albada}}{{Roberts} et~al.}{1979}]{Roberts1979}
{Roberts} Jr. W.~W.,  {Huntley} J.~M.,   {van Albada} G.~D.,  1979, \mn@doi [\apj] {10.1086/157367}, \href {https://ui.adsabs.harvard.edu/abs/1979ApJ...233...67R} {233, 67}

\bibitem[\protect\citeauthoryear{{Savchenko} \& {Reshetnikov}}{{Savchenko} \& {Reshetnikov}}{2013}]{Savchenko2013}
{Savchenko} S.~S.,  {Reshetnikov} V.~P.,  2013, \mn@doi [\mnras] {10.1093/mnras/stt1627}, \href {https://ui.adsabs.harvard.edu/abs/2013MNRAS.436.1074S} {436, 1074}

\bibitem[\protect\citeauthoryear{{Sellwood} \& {Binney}}{{Sellwood} \& {Binney}}{2002}]{Sellwood2002}
{Sellwood} J.~A.,  {Binney} J.~J.,  2002, \mn@doi [\mnras] {10.1046/j.1365-8711.2002.05806.x}, \href {https://ui.adsabs.harvard.edu/abs/2002MNRAS.336..785S} {336, 785}

\bibitem[\protect\citeauthoryear{{Sellwood} \& {Carlberg}}{{Sellwood} \& {Carlberg}}{2019}]{Sellwood2019}
{Sellwood} J.~A.,  {Carlberg} R.~G.,  2019, \mn@doi [\mnras] {10.1093/mnras/stz2132}, \href {https://ui.adsabs.harvard.edu/abs/2019MNRAS.489..116S} {489, 116}

\bibitem[\protect\citeauthoryear{{Shu}}{{Shu}}{1970}]{Shu1970}
{Shu} F.~H.,  1970, \mn@doi [\apj] {10.1086/150410}, \href {https://ui.adsabs.harvard.edu/abs/1970ApJ...160...99S} {160, 99}

\bibitem[\protect\citeauthoryear{{Toomre}}{{Toomre}}{1964}]{Toomre1964}
{Toomre} A.,  1964, \mn@doi [\apj] {10.1086/147861}, \href {https://ui.adsabs.harvard.edu/abs/1964ApJ...139.1217T} {139, 1217}

\bibitem[\protect\citeauthoryear{{Toomre}}{{Toomre}}{1981}]{Toomre1981}
{Toomre} A.,  1981, in {Fall} S.~M.,  {Lynden-Bell} D.,  eds, Structure and Evolution of Normal Galaxies. pp 111--136

\bibitem[\protect\citeauthoryear{{Tremaine}}{{Tremaine}}{2023}]{Tremaine2023}
{Tremaine} S.,  2023, {Dynamics of Planetary Systems}.
Princeton University Press, Princeton NJ

\bibitem[\protect\citeauthoryear{{Zhang}}{{Zhang}}{1996}]{Zhang1996}
{Zhang} X.,  1996, \mn@doi [\apj] {10.1086/176717}, \href {https://ui.adsabs.harvard.edu/abs/1996ApJ...457..125Z} {457, 125}

\bibitem[\protect\citeauthoryear{{Zhang} \& {Buta}}{{Zhang} \& {Buta}}{2007}]{Zhang2007}
{Zhang} X.,  {Buta} R.~J.,  2007, \mn@doi [\aj] {10.1086/514337}, \href {https://ui.adsabs.harvard.edu/abs/2007AJ....133.2584Z} {133, 2584}

\makeatother
\end{thebibliography}

\begin{appendix}




\onecolumn

\section{Proof of equation~\eqref{EQ:SURF:FROM:PSI}}
\label{app:proof}
I begin with the observation that the potential of a razor-thin disc can always be expressed as $\Phi(\vec{r})=F(\vec{R},|z|)$, where $F(\vec{r})$ is a smooth function satisfying $\laplacian F =0$ for $z\ge0$ (one can construct $F$ by setting $F=\Phi$ for $z\ge0$ and by analytic continuation for $z < 0$). Then $\Psi(\vec{R}) = F (\vec{R}, 0)$ and Poisson’s equation at $z = 0$ implies
\begin{align}
    \label{eq:F:Sigma}
    2\pi G \Sigma(\vec{R}) &= \lim_{z\to0^+} \pdv{\Phi}{z} = \eval{\pdv{F}{z}}_{z=0} , \qquad\text{and} \\
    \label{eq:F:Psi}
    \laplacian\Psi &= -\eval{\pdv[2]{F}{z}}_{z=0}.
\end{align}
Because $\tilde{F}(\vec{r}) \equiv -\partial F/\partial z$ also satisfies $\laplacian\tilde{F} = 0$, another razor-thin model can be constructed, whose potential in the plane
\begin{align}
    \label{eq:tF:tPsi}
    \tilde\Psi(\vec{R})=\tilde{F}(\vec{R},0)= -\eval{\pdv{F}{z}}_{z=0}=-2\pi G\Sigma(\vec{R}).
\end{align}
For the surface density of the new model, Eq.~\eqref{eq:F:Sigma} applied to $\tilde{F}$ gives
\begin{align}
    \label{eq:tF:tSigma}
    2\pi G\tilde{\Sigma}(\vec{R}) &= \eval{\pdv{\tilde{F}}{z}}_{z=0} = -\eval{\pdv[2]{F}{z}}_{z=0} = \laplacian \Psi,
\end{align}
where the last equality follows from Eq.~\eqref{eq:F:Psi}. Finally, when the Poisson integral~\eqref{eq:Psi:from:Surf} relating $\tilde{\Psi}$ and $\tilde{\Sigma}$ is expressed in terms of $\Sigma$ and $\laplacian\Psi$ using Eqs.~(\ref{eq:tF:tPsi}, \ref{eq:tF:tSigma}), Eq.~\eqref{EQ:SURF:FROM:PSI} follows. \qed

\section{Evaluating the Poisson integral numerically}
\label{app:Poisson:numerics}
Assume that the density has been decomposed into azimuthal Fourier components
\begin{align}
    \Sigma(\vec{R}) &= \sum_{m=-\infty}^\infty \Sigma_m(R) \,\Exp{\imag m\phi}
    \qq{with}
    \Sigma_m(R) = \frac1{2\pi}\int_0^{2\pi}\mathrm{d}\phi\,\Sigma(\vec{R})\,\Exp{-\imag m\phi},
\end{align}
where the $\Sigma_m(R)$ are generally complex-valued and often given in polar form $\Sigma_m(R)=S_{\!m}(R)\,\Exp{-\imag m\psi(R)}$ with real-valued amplitude $S_{\!m}$ and phase $\psi$. When this is inserted into the Poisson integral~\eqref{eq:Psi:from:Surf}, the azimuthal Fourier component of the potential is found to be
\begin{align}
    \label{eq:Psi:m:from:Surf:m}
    \Psi_m(R) &= -G \int_0^\infty \dd R'\,R'\,\Sigma_m(R') \int_0^{2\pi}\frac{\dd\phi'\,\Exp{\imag m(\phi'-\phi)}}{\sqrt{R^2-2RR'\cos|\phi-\phi'|+R'^2}}
    && =-\pi G \int_0^\infty\mathrm{d}R'\,\Sigma_m(R')\; b_{1/2}^m\!\left(R/R'\right),
\end{align}
where 
\begin{align}
    \label{eq:Poisson:Laplace}
    b_{s}^m(\alpha)
    = \frac1\pi\int_0^{2\pi}\frac{\Exp{\imag m\varphi}\,\dd\varphi}{\left(1-2\alpha\cos\varphi+\alpha^2\right)^s}
    = \frac1{2\pi}\int_0^{\pi}\frac{\cos m\varphi\,\dd\varphi}{\left(1-2\alpha\cos\varphi+\alpha^2\right)^s}
\end{align}
are the `Laplace coefficients'. Their computation is detailed in Appendix~\ref{app:laplace}, while Fig.~\ref{fig:LaplaceCoeffs} plots them for $s=\sfrac12$. At $\alpha=1$, $b_{1/2}^m$ has a logarithmic singularity for all $m$. Other important properties are
\begin{align}
    \label{eq:Laplace:props}
    b_s^{-m}(\alpha) &=b_s^m(\alpha), &
    b_s^m(\alpha^{-1}) &=\alpha^{2s}b_s^m(\alpha), &
    b_s^{m}(\alpha) &\sim\alpha^m\;\text{as $\alpha\to0$,}\qand &
    b_s^{m}(\alpha) &\sim\alpha^{-2s-m}\; \text{as $\alpha\to\infty$}.
\end{align}

In order to compute the derivative $\dd\Psi_m/\dd R$ needed for the force, the derivative of $b_{1/2}^m$ in Eq.~\eqref{eq:Psi:m:from:Surf:m} could be taken via the relation \citep[][eq.~7.4.13]{CeMe1972}
\begin{align}
    \frac{\dd b_s^m}{\dd\alpha} &= \frac{[m+(m+2s)\alpha^2]b_s^m(\alpha)-2[m+1-s]b_s^{m+1}(\alpha)}{\alpha(1-\alpha^2)}.
\end{align}
Alternatively, because Eq.~\eqref{eq:Psi:m:from:Surf:m} is a convolution, the derivative can be shifted onto $\Sigma_m$:
\begin{align}
    \label{eq:dPsi:m:from:dSurf:m}
    \frac{\dd\Psi_m}{\dd R} &= -\frac{\pi G}R \int_0^\infty \dd R' \,\frac{\dd (R'\Sigma_m)}{\dd R'}\,b_{1/2}^m(R/R').
\end{align}

\begin{figure}
\begin{center}
 	\includegraphics[width=0.9\linewidth]{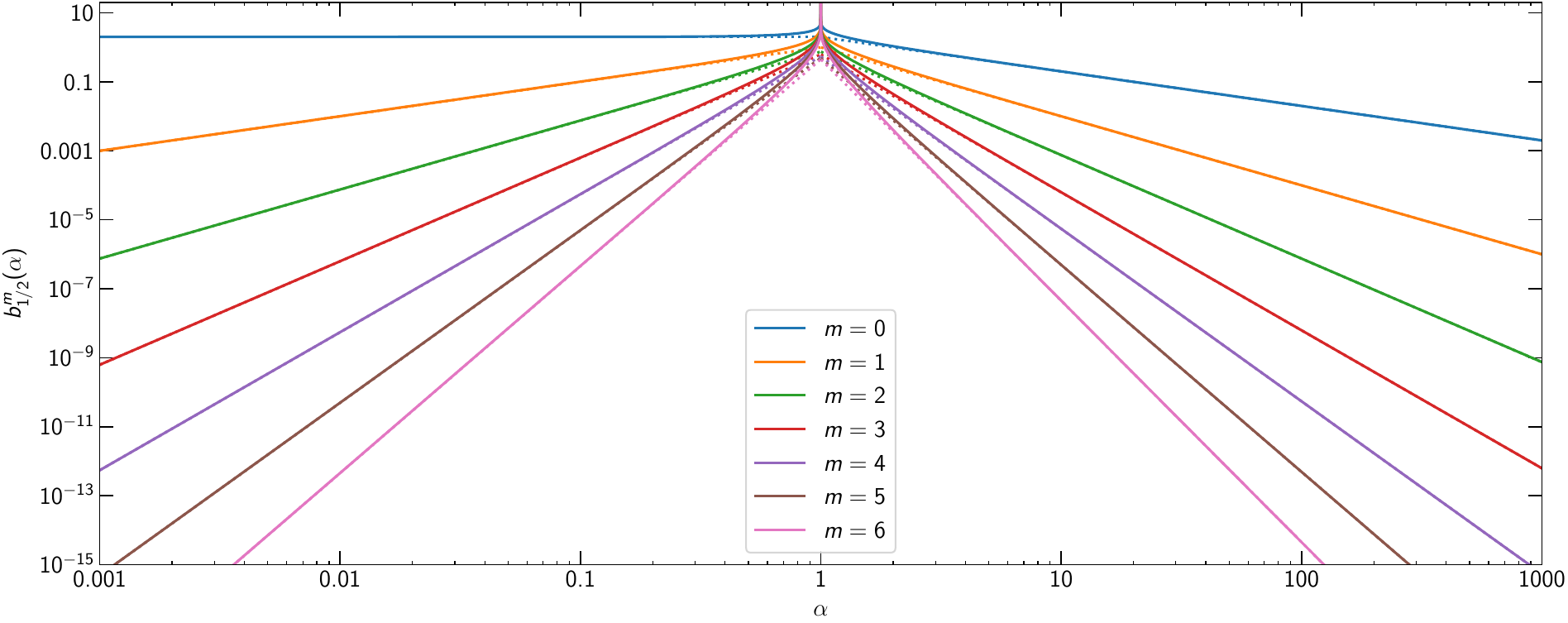}
    \caption{
    The Laplace coefficients $b_{1/2}^m(\alpha)$ (solid) and their asymptotes at $\alpha\ll1$ and $\alpha\gg1$ (dotted).}
    \label{fig:LaplaceCoeffs}
    \vspace*{-1mm}   
\end{center}
\end{figure}

My aim is to employ Gau\ss{}-Legendre quadrature to evaluate the integrals in~\eqref{eq:Psi:m:from:Surf:m} and~\eqref{eq:dPsi:m:from:dSurf:m}. This requires to transform to integrals with finite boundaries and finite integrand. To avoid the infinite integration interval and shift the singularity to $x=0$, I use the substitution
\begin{align}
    x = \frac{R'-R}{R'+R+a},
    \qquad R' = \frac{R+x(R+a)}{1-x}
\end{align}
with some scale length $a\ge0$ (for small $R$ it is often better to set $a>0$), such that the integral~\eqref{eq:Psi:m:from:Surf:m} becomes
\begin{align}
    \Psi_m(R) &=\pi G \int_{-R/(R+a)}^1 \mathrm{d}x \,\frac{2R+a}{(1-x)^2}\, \Sigma_m\big(R'(x)\big)\,b_{1/2}^m\big(R/R'(x)\big).
\end{align}
The logarithmic singularity at $x=0$ can now be dealt with by the second substitution $x=u^3$ or $x=u^5$, such that the integrand is smooth at $u=0$. Alternatively, the singularity can be treated by a technique used by \cite{Hure2005} in a very similar context. Suppose we already know a potential-density pair $\bar\Psi_m(R)$, $\bar\Sigma_m(R)$ that is an exact solution. Then
\begin{subequations}
    \begin{align}
    \Psi_m(R) = \Sigma_m(R) \frac{\bar\Psi_m(R)}{\bar\Sigma_m(R)} -\pi G\int_0^\infty\dd R' \left[\Sigma_m(R')-\Sigma_m(R)\frac{\bar\Sigma_m(R')}{\bar\Sigma_m(R)}\right]\,b_{1/2}^m(R/R').
\end{align}
Thus, the integral only computes the difference between $\Psi_m$ and $(\Sigma_m/\bar\Sigma_m)$ times $\bar\Psi_m$. The point of this construct is that the integrand now vanishes at $R'=R$ and the previous method without the second substitution can be used. A suitable choice for $\bar\Psi_m$ and $\bar\Sigma_m$ is the model~\eqref{eqs:Kuzmin:m} with parameter $a$ set such that $\bar\Sigma_m(R')$ is maximal at $R'=R$, giving
\begin{align}
    \frac{\bar\Psi_m(R)}{\bar\Sigma_m(R)} &= -\frac{2\pi GR}{\sqrt{m(m+3)}}\frac{2m+3}{2m+1},&
    \frac{\bar\Sigma_m(R')}{\bar\Sigma_m(R)} &=\frac{(2m+3)^{m+3/2} (R'/R)^m}{(m+3+m [R'/R]^2)^{m+3/2}}.
\end{align}
\end{subequations}

\section{Computing the Laplace coefficients}
\label{app:laplace}
The Laplace coefficients $b_s^m(\alpha)$, defined in Eq.~\eqref{eq:Poisson:Laplace}, have been studied by Laplace, Lagrange, Legendre, Euler, Cauchy, and Gauß amongst many others and are covered in every textbook on celestial mechanics \citep[e.g.][]{CeMe1972,Tremaine2023}. Because of their properties~\eqref{eq:Laplace:props}, their computation can be limited to $m\ge0$ and $0\le\alpha\le1$, and I assume that these conditions are satisfied. The Laplace coefficients are related to the associated Legendre functions of the second kind, $Q_\nu^\mu$, with half-integer degree $\nu$ and integer order $\mu$ (also known as toroidal functions) via
\begin{align}
    \label{eq:b:Q}
    b_s^m(\alpha) &= \frac2{\Gamma(s)\sqrt{\pi\alpha}(\alpha^2-1)^{s-1/2}} \,Q_{m-1/2}^{s-1/2}(\chi)
    \qquad\text{for $\alpha<1$}\qquad\text{with}\quad\chi \equiv \tfrac12(\alpha+\alpha^{-1}).
\end{align}
A representation due to Lagrange \citep[][eq.~7.1.2]{CeMe1972} is
\begin{subequations}
    \label{eqs:b:F21}
\begin{align}
    \label{eq:b:F21:def}
    b_s^m(\alpha) &= \frac{2\Gamma(m+s)}{\Gamma(s)\Gamma(m+1)}\,\alpha^{m}\,{}_2F_1\left(s,m+s;\,m+1;\,\alpha^2\right)
    =
    \frac{2\alpha^m}{\Gamma(s)^2} \sum_{n=0}^\infty\frac{\Gamma(n+s)\Gamma(m+n+s)}{\Gamma(n+1)\Gamma(m+n+1)}\alpha^{2n},
\end{align}
where ${}_2F_1$ is the hypergeometric function. One possibility to compute the Laplace coefficients is via the hypergeometric series (the second expression in Eq.~\ref{eq:b:F21:def}), which gives the algorithm
\begin{align}
    \label{eq:B:F21:algo}
    b_s^m(\alpha) &= \sum_{n=0}^\infty t_{s,n}^{m}, &
    t_{s,0}^{m} &= 2\frac{(s)_m}{m!}\alpha^m,&
    \frac{t_{s,n+1}^{m}}{t_{s,n}^{m}} &= \frac{(n+s)(m+n+s)}{(n+1)(m+n+1)} \alpha^2,
\end{align}
\end{subequations}
where $(a)_n=a(a+1)\dots(a+n-1)=\Gamma(a+n)/\Gamma(a)$ is the (rising) Pochhammer symbol. This algorithm converges for $\alpha<1$, but only very slowly for $\alpha\to1$, where $b_s^m(\alpha)$ is singular. 
\begin{subequations}
For $s=\sfrac12$, the singularity at $\alpha=1$ is logarithmic, i.e.\ $b_{1/2}^m(\alpha) \sim \ln(1-\alpha)$, and a suitable alternative expression based on Eq.~(15.3.10) of \cite{AS} is
\begin{align}
    \label{eq:b:F21:pole}
    b_{1/2}^m(\alpha) 
    &= \frac{2\alpha^m}{\pi^{3/2}\Gamma(m+\tfrac12)} \sum_{n=0}^\infty \frac{\Gamma(n+\tfrac12)\Gamma(m+n+\tfrac12)}{n!^2}\left[2\psi(n+1)-\psi(n+\tfrac12)-\psi(m+n+\tfrac12)-\ln\epsilon\right]\epsilon^n,
\end{align}
where $\epsilon=1-\alpha^2$ and $\psi(z)\equiv\Gamma'(z)/\Gamma(z)$ is the digamma function. The resulting algorithm is
\begin{align}
    \label{eq:b:F21:pole:algo}
    b_{1/2}^m(\alpha)
    &=\sum_{n=0}^\infty p_n^m\, q_n^m, &
    p_0^m &= \frac{2\alpha^m}{\pi}, &\frac{p_{n+1}^m}{p_n^m} &=  \frac{(n+\frac12)(m+n+\frac12)}{n^2} \epsilon^2, \\
    &&q_0^m &= 4\ln2 - \ln\epsilon - \sum_{k=0}^{m-1}\frac1{k+\frac12} ,&q_{n+1}^m&= q_n^m + \frac2{n+1} - \frac1{n+\frac12} - \frac1{m+n+\frac12}. \nonumber
\end{align}
\end{subequations}

\begin{figure}
	\includegraphics[width=0.5\linewidth]{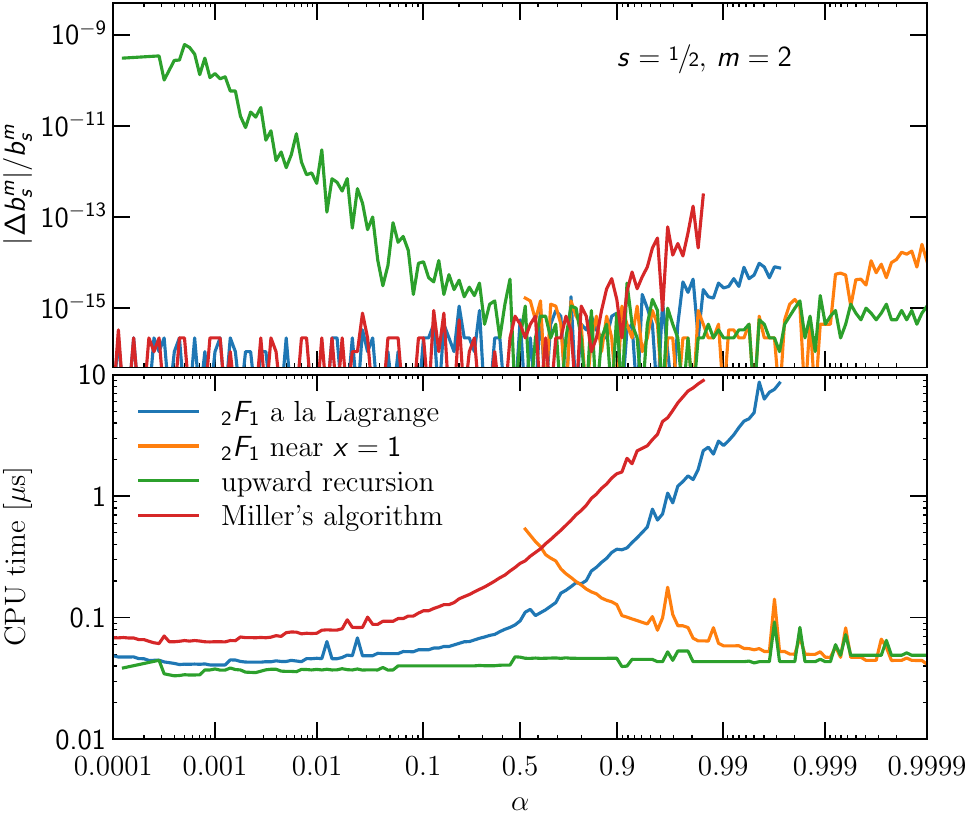}
    \hfill
	\includegraphics[width=0.5\linewidth]{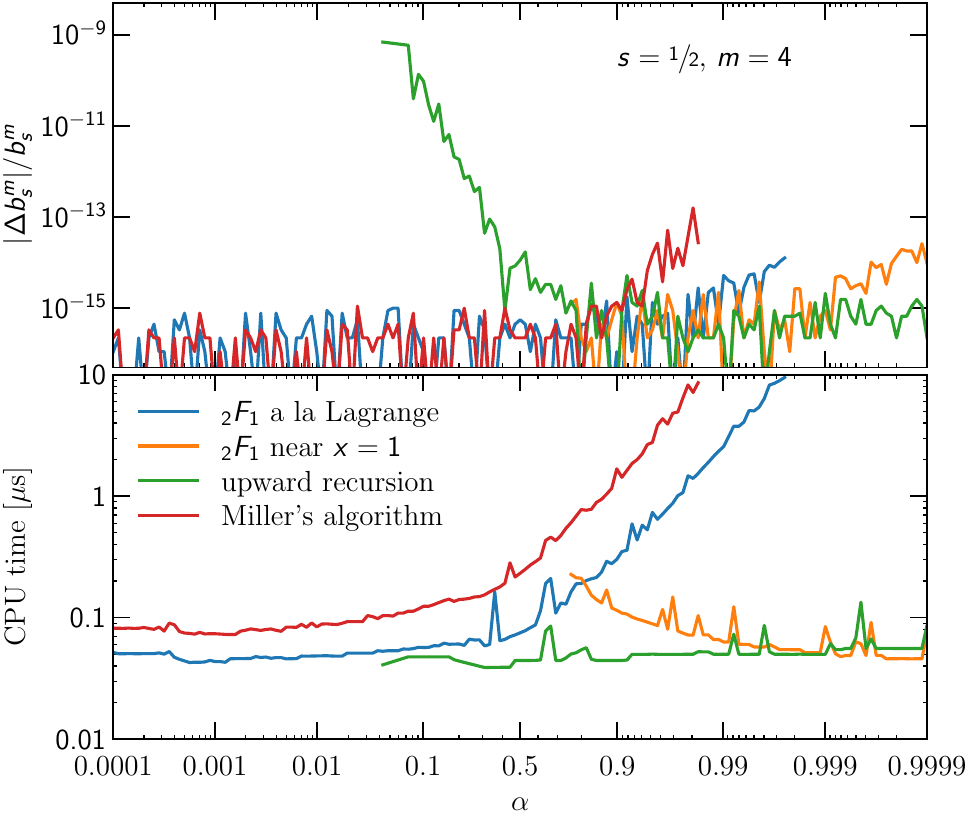}
    \caption{
    Relative error (\textbf{top}) and costs (\textbf{bottom}) of the computation of the Laplace coefficient $b_s^m$ via four different methods described in the text for $s=\sfrac12$ and $m=2$ (\textbf{left}) or $m=4$ (\textbf{right}).}
    \label{fig:LaplaceCoeff:compute}
    \vspace*{2mm}
\end{figure}

Potentially more efficient algorithms are based on the recursion relations
\begin{subequations}
\begin{align}
    \label{eq:b:recur}
    (m+1-s) \,b_s^{m+1}(\alpha) &= m \left[\alpha+\alpha^{-1}\right] \,b_s^{m}(\alpha) - (m-1+s) \,b_s^{m-1}(\alpha)
\end{align}
\citep[Euler in][eq 7.7]{CeMe1972} to increment or decrement $m$ and 
\begin{align}
    s(1\pm\alpha)^2\left[b_{s+1}^m(\alpha)\mp b_{s+1}^{m+1}(\alpha)\right] &= (m+s)b_s^m(\alpha)\pm (m-s+1)b_s^{m+1}(\alpha)
\end{align}
\end{subequations}
\citep[][eq.~7.11]{CeMe1972} to increment or decrement $s$. From these and some starting values for $b_s^{m}$ and $b_s^{m+1}$, the values of the Laplace coefficients for any $s$ and $m$ can, in theory, be computed. Possible starting values are $b_{1/2}^0(\alpha) =  4 K(\alpha)/\pi$ and $b_{1/2}^1(\alpha) =  4[K(\alpha)-E(\alpha)]/\pi\alpha$, where $K$ and $E$ are complete elliptical integrals of the first and second kind, respectively. These can be efficiently computed simultaneously from their relation to Gauß' arithmetic-geometric mean (\citealt{NIST-DLMF}, \S\href{https://dlmf.nist.gov/19.8i}{19.8i}), which gives the algorithm
\begin{subequations}
\label{eqs:b0,1:algo}
\begin{align}
    \label{eqs:b0:algo}
    a_0,\;g_0 &= 1,\,\sqrt{1-\alpha^2}, & a_{n+1},\;g_{n+1} &= \tfrac12(a_n+g_n),\;\sqrt{a_ng_n}, & b_{1/2}^0(\alpha) &= 2/a_\infty, \\
    \label{eqs:b1:algo}
    c_0 &= \tfrac12 \alpha^2, & c_{n+1} &= c_n + 2^{n-3}\sqrt{a_n-g_n}, &
    b_{1/2}^1(\alpha) &= b_{1/2}^0(\alpha)\, c_\infty/\alpha,
\end{align}
\end{subequations}
which converges quadratically to $a_\infty=g_\infty$.

Unfortunately, the recursion~\eqref{eq:b:recur} is unstable in the upwards direction\footnote{The recursion~\eqref{eq:b:recur} has two solutions, one dominant (growing) and the other recessive (shrinking) as $m$ is incremented. The $b_s^m$ are the recessive solution, while the dominant solution is obtained by replacing the associated Legendre functions of the second kind in Eq.~\eqref{eq:b:Q} with those of the first kind. Any small numerical error of $b_s^0$ and $b_s^1$ corresponds to this dominant solution, which will dominate the numerical result for $b_s^m$ at $m\gg1$.} and hence cannot be used to compute the Laplace coefficients for $\alpha^m\ll1$ with sufficient accuracy when starting from $m=0,1$. In that regime, one may either use the series expression (algorithm~\ref{eq:B:F21:algo}) or apply the recursion~\eqref{eq:b:recur} downwards (Miller's algorithm). To this end, one starts from some value $m'\gg m$ with initial values $b_s^{m'+1}=\alpha$ and $b_s^{m'}=1$, corresponding to the asymptotic behaviour $b_s^{m+1}/b_s^m\sim\alpha$ at small $\alpha$. Next, the recursion is applied to obtain corresponding values for $b_s^{m'-1},\,b_s^{m'-2}, \dots, b_s^{0}$. At small $m$, these are only wrong by the same overall normalisation (for the same reason that the recursion is unstable in the upwards direction), which for $s=\sfrac12$ can be obtained from $b_{1/2}^0(x)$ via~\eqref{eqs:b0:algo} and in general from the relation \cite[][problem 4.6]{Tremaine2023}
\begin{align}
    \label{eq:b:sum}
    \tfrac12b_s^0(\alpha) + \sum_{m=1}^\infty b_s^m(\alpha) &= (1-\alpha)^{-2s}.
\end{align}
The asymptotic $b_s^m(\alpha)\sim\alpha^{m}$ for small $\alpha$ implies that reaching a precision $\epsilon$ requires $\alpha^{m'-m}=\epsilon$ or $m'=m+\ln\epsilon/\ln\alpha=m-16/\log_{10}(\alpha)$ for the usual 64-bit floating-point arithmetic. These recursive methods are particularly useful for computing several Laplace coefficients with different $s$ or $m$, but even when only a single coefficient is required, as in our application, they are competitive.

This gives four different methods to compute the Laplace coefficients. For $s=\sfrac12$, the only value actually required by the methods of Appendix~\ref{app:Poisson:numerics}, I assessed them by comparison to a calculation based on the first equality in~\eqref{eq:b:F21:def} computed using implementations for the hypergeometric and Gamma functions from the Gnu Scientific Library  \citep[\texttt{gsl},][]{GSL}. At $\alpha\to1$ and $m\gg1$, the \texttt{gsl} routines become inaccurate or even fail altogether (exit with an error), in which case I took the mean of the two least deviating values from my four methods as estimate for the correct value. Fig.~\ref{fig:LaplaceCoeff:compute} plots the resulting relative errors (top) and the required CPU time (bottom) for $b_{1/2}^2$ (left) and $b_{1/2}^4$ (right). It appears that for small $\alpha$ the direct summation of the hypergeometric series (blue) is accurate and most efficient, while at larger $\alpha$ the upward recursion (green) is the best method.

\section{The potential of the scale-invariant spirals}
\label{app:K}
Following the ansatz of Appendix~\ref{app:proof}, the potential can be expressed as $\Phi(\vec{r})=F(\vec{R},|z|)$ with a smooth function $F(\vec{r})$ that satisfies
\begin{align}
    \label{eq:cond:F}
    \laplacian F&=0\quad\text{at $z\ge0$}, \quad
    F\to0\quad\text{as $z\to\infty$},\quad\text{and}\quad
    \eval{\pdv{F}{z}}_{z=0} = 2\pi G\Sigma(\vec{R}).
\end{align}
Since $\Sigma(\vec{R})= S_{\!m,0}(R/R_0)^{-\gamma-\imag m\lambda} \Exp{\imag m\phi} \propto R^{-3/2-\imag \zeta} \Exp{\imag m\phi}$ is scale-invariant, a natural ansatz for $F(\vec{r})$ is a scale-invariant form, too:
\begin{align}
    F(\vec{r}) &= -2\pi G\,R_0 \, S_{\!m,0}\, (r/R_0)^{-1/2-\imag\zeta}\,f_m(\cos\theta)\,\Exp{\imag m\phi},
\end{align}
with spherical coordinates $(r,\theta,\phi)$. The function $f_m$ must be determined by the conditions~\eqref{eq:cond:F}, in particular (with $x=\cos\theta$)
\begin{align}
    \label{eq:Laplace:F}
    \laplacian F \propto (r/R_0)^{-5/2-\imag\zeta}\,\Exp{\imag m\phi} \left\{\left[\qty(-\tfrac12-\imag\zeta)\qty(\tfrac12-\imag\zeta) -\frac{m^2}{1-x^2}\right] f_m(x) + \dv{x}(1-x^2)\dv{f_m}{x}\right\} = 0.
\end{align}
For this to hold and $f\to0$ as $x\to1$ (so that $F\to0$ as $z\to\infty$), $f_m(x) = C P^{\,|m|}_{-1/2-\imag\zeta}(x)$, where $P_\nu^{\,\mu}$ is the associated Legendre function of the first kind. The constant $C$ is determined by the last of the conditions~\eqref{eq:cond:F}, giving $C^{-1}=-\dd P^{\,|m|}_{-1/2-\imag\zeta}/\dd x\big|_{x=0}$. Thus,
\begin{align}
    \Phi(\vec{r}) &= -2\pi G\, R_0 S_{\!m,0}\, f_m(|\cos\theta|)\,(r/R_0)^{1-\gamma}\, \Exp{\imag m[\phi-\lambda\ln(r/R_0)]}, &
    \label{eq:Km:def}
    f_m(x) &= -\qty(\dd P^{\,|m|}_{-1/2-\imag\zeta}/\dd x)^{-1}_{x=0}\, P^{\,|m|}_{-1/2-\imag\zeta}(x).
\end{align}
For real-valued $\zeta$ (at $\gamma=\sfrac32$), the product $(-\sfrac12-\imag\zeta)(\sfrac12-\imag\zeta)$ appearing in Eq.~\eqref{eq:Laplace:F} is also real and, consequently, so is $P_\nu^{\,m}(z)$ for this particular case (and known as `conical function'). 

At $z=0$, $\cos\theta=0$ and
\begin{align}
    \label{EQ:K(0)}
    K_m(\zeta) \equiv f_m(0) = -\qty(\dv{\ln P^{\,|m|}_{-1/2-\imag\zeta}}{x})^{-1}_{x=0} &=
    \frac12 \frac{\Gamma\left(\tfrac14-\tfrac12m-\tfrac\imag2\zeta\right)\,\Gamma\left(\tfrac14-\tfrac12m+\tfrac\imag2\zeta\right)}{\Gamma\left(\frac34-\frac12m-\frac\imag2\zeta\right)\,\Gamma\left(\frac34-\frac12m+\frac\imag2\zeta\right)}
    = \frac12 \frac{\Gamma\left(\tfrac14+\tfrac12m-\tfrac\imag2\zeta\right)\,\Gamma\left(\tfrac14+\tfrac12m+\tfrac\imag2\zeta\right)}{\Gamma\left(\frac34+\frac12m-\frac\imag2\zeta\right)\,\Gamma\left(\frac34+\frac12m+\frac\imag2\zeta\right)},
\end{align} 
where I have used Eqs.~\href{https://dlmf.nist.gov/14.5.E1}{14.5.1},\href{https://dlmf.nist.gov/14.5.E2}{2} of \cite{NIST-DLMF} and applied $\Gamma(z)\Gamma(1-z)=\pi/\sin\pi z$ to each Gamma factor for the last equality.

\subsection{Alternative derivations of Eq.~\eqref{EQ:K(0)}}
Eq.~\eqref{EQ:K(0)} is merely an extension of \citeauthor{Kalnajs1971}' (\citeyear{Kalnajs1971}) Eq.~(12) to complex-valued $\zeta$. While \cite{Kalnajs1971} did not provide a derivation, he later\footnote{Juli 2000, in a letter to Howard Cohl, privately communicated to the author in 2025.} sketched a proof as follows. Inserting the surface density~\eqref{eq:Sigma:power-law} into Eq.~\eqref{eq:Psi:m:from:Surf:m} gives the integral expression for $K_m(\zeta)$ in Eq.~\eqref{eq:T:S*K}. Inserting the series~\eqref{eq:b:F21:def} for $b_{1/2}^m(\Exp{-|w|})$ and evaluating the integrals over $w$ directly gives the pole expansion for the meromorphic function $K_m(\zeta)$, which has simple poles at $\zeta=\pm\imag(\sfrac12+m+2n)$ for $n\in\mathbb{N}$. It is straightforward to show that Eq.~\eqref{EQ:K(0)} has the same simple poles and residues and hence the same pole expansion. Thus, according to Mittag-Leffler's theorem, $K_m(\zeta)$ can differ from the form~\eqref{EQ:K(0)} at most by a constant, but as both vanish at infinity, this constant is zero and the two functions identical.

Following a method due to \citeauthor{Tremaine2023} (\citeyear{Tremaine2023}, Box~6.2), yet another way to derive Eq.~\eqref{EQ:K(0)} constructs the scale-invariant spirals as continuous distributions of the models~\eqref{eqs:Kuzmin:m} with different scale lengths $a$, normalisations $v_0^2(a)\propto a^{-\gamma}$, and phases $\psi(a)=\lambda\ln a$, or combined, with complex-valued $v_0^2(a)\propto a^{-3/2-\imag\zeta}$:
\begin{alignat}{3}
    \label{eq:Sigma:dist}
    \Sigma(\vec{R}) &\propto
    \frac{2m+1}{2\pi} \Exp{\imag m\phi}
    \int_0^\infty \frac{R^ma^{m+1/2-\imag\zeta}\,\dd a}{(a^2+R^2)^{m+3/2}}
    &&=\frac{2m+1}{4\pi}\, 
    B\qty(\tfrac34+\tfrac{m}2+\imag\tfrac\zeta2,\tfrac34+\tfrac{m}2-\imag\tfrac\zeta2)\, R^{-3/2-\imag\zeta}\,\Exp{\imag m\phi}
    , \\
    \label{eq:Psi:dist}
    \Psi(\vec{R}) &\propto -G \Exp{\imag m\phi} \int_0^\infty \frac{R^ma^{m-1/2-\imag\zeta}\,\dd a}{(a^2+R^2)^{m+1/2}}
    &&= -\frac{G}{2}\, 
    B\qty(\tfrac14+\tfrac{m}2+\imag\tfrac\zeta2,\tfrac14+\tfrac{m}2-\imag\tfrac\zeta2)\, R^{-1/2-\imag\zeta}\, \Exp{\imag m\phi}.
\end{alignat}
Here, I have used $\int_0^\infty t^{a-1}(1+t)^{-a-b}\dd t=B(a,b)=\Gamma(a)\Gamma(b)/\Gamma(a+b)$, which requires the real parts for $a$ and $b$ to be positive, which in turn implies $|\gamma-\sfrac32|<m+\sfrac12$ and constitutes a limitation of this derivation. Because $R^{-3/2-\imag\zeta}=R^{-\gamma}\,\Exp{-\imag m\lambda\ln R}$, $\Sigma(\vec{R})$ in Eq.~\eqref{eq:Sigma:dist} is proportional to that of the scale-invariant models, and eliminating the same constant of proportionality between Eqs.~(\ref{eq:Sigma:dist}, \ref{eq:Psi:dist}) gives Eq.~\eqref{EQ:K(0)}.

\subsection{Computing the $P_\nu^{\,m}$}
\label{app:scale-free:P}
Unfortunately, standard libraries (e.g.\ \texttt{gsl} or \texttt{scipy}) do not currently support the computation of the Legendre functions for complex degree $\nu$ (needed for the potential shape function $f_m(x)$), but its computation is straightforward. For integer $m$
\begin{align}
    \label{eq:Pmnu}
    P_\nu^{\,m}(z) &= \frac{(-\nu)_m(\nu+1)_m}{\Gamma(m+1)} \left[\frac{1-z}{1+z}\right]^{m/2}{}_2F_1\left(-\nu,\nu+1;m+1;\tfrac12(1-z)\right)
    = \frac{(-\nu)_m(\nu+1)_m}{\Gamma(m+1)} \left[\frac{1-z}{1+z}\right]^{m/2}
    \sum_{n=0}^\infty \frac{(-\nu)_n(\nu+1)_n}{(m+1)_n\, n!} \qty(\frac{1-z}2)^n
\end{align}
(e.g.~Eq.~\href{https://dlmf.nist.gov/14.3.E5}{14.3.5} of \citealt{NIST-DLMF}), which can be computed analogously to algorithm~\eqref{eqs:b:F21} for the computation of the Laplace coefficient.

\section{The gravitational torque}
\label{app:torque}
When the spiral density perturbation $\Sigma(\vec{R}) = S_{\!m}(R)\Exp{\imag m(\phi-\psi)}$ (as in Eq.~\ref{eq:spiral:Surf}) is inserted into Eq.~\eqref{eq:Psi:m:from:Surf:m}, its gravitational potential is obtained as
\begin{align}
    \label{eq:Pm:from:Surf}
    \Psi(  \vec{R}) =-P_{m}(R)\Exp{\imag m(\phi-\psi)}
    \qq{with}
    P_{\!m}(R) = \pi G \int_0^\infty\dd R'\;S_{\!m}(R')\, \Exp{\imag m(\psi-\psi')}\, b_{1/2}^m(R/R'),
\end{align}
where $\psi=\psi(R)$ and $\psi'=\psi(R')$. While $S_{\!m}$ is real-valued by definition, $P_{\!m}(R)$ is in general complex-valued (differently from our convention in Eq.~\ref{eq:spiral:Psi}). Therefore (and because only the real parts of $\Sigma(\vec{R})$ are $\Psi(\vec{R})$ have meaning), the torque density is
\begin{align}
    \tau = -\Sigma(\vec{R}) \,\pdv{\Psi}{\phi}
    \quad = -m S_{\!m} \cos\qty(m\qty[\phi-\psi]) \Big[\Re\qty{P_{\!m}}\sin\qty(m\qty[\phi-\psi]) + \Im\qty{P_{\!m}}\cos\qty(m\qty[\phi-\psi])\Big].
\end{align}
After integrating over $\phi$, only the term involving the imaginary part of $P_{\!m}$ remains and one finds for the total torque (using Eq.~\ref{eq:Pm:from:Surf})
\begin{align}
    \label{eq:torque:cyl}
    T = -m\pi \int_0^\infty\dd R\,R\,S_{\!m}\Im\{P_{\!m}\}
    \quad = -m\pi^2 G 
    \int_0^\infty\dd R\int_0^\infty\dd R'\, S_{\!m}(R)\, S_{\!m}(R')\, \sin \big(m[\psi-\psi']\big)\,R\,b_{1/2}^m(R/R') = 0.
\end{align}
Since $Rb_{1/2}^m(R/R')=R'b_{1/2}^m(R'/R)$, the integrand is anti-symmetric with respect to swapping $R$ and $R'$ and the integral vanishes. When $S_{\!m}(R)=S_{\!m,0}\,\Exp{-\gamma u}$ and $\psi(R)=\lambda u$ with $u\equiv\ln(R/R_0)$ for the scale-invariant spirals of Section~\ref{sec:spiral:scale-invariant} is inserted into Eq.~\eqref{eq:torque:cyl} and the integration variables are changed to $u$ and $u'$, it becomes
\begin{subequations}
\begin{align}
    T &= -m \pi^2 G R_0^3S_{\!m,0}^2 \int_{-\infty}^\infty \dd u \int_{-\infty}^\infty \dd u'\, \sin\big(m\lambda[u-u']\big)\,\Exp{(\frac32-\gamma)(u'+u)-\frac12|u'-u|}\, b_{1/2}^m\Big(\Exp{-|u'-u|}\Big) = 0,
\end{align}
where I used $R b_{1/2}^m(R/R') = R' b_{1/2}^m(R'/R) = R_0 \Exp{\frac12(u'+u-|u'-u|)}b_{1/2}^m(\Exp{-|u'-u|})$. The integrand is, of course, still anti-symmetric with respect to swapping $u$ with $u'$, such that the total torque vanishes, as it should. Factorising the double integral into the product of two one-dimensional integrals by changing the inner integration variable from $u'$ to $w=u'-u$ gives
\begin{align}
    \label{eq:T:S*K}
    T &=
    -m 2 \pi^2 G \,\underbrace{R_0^3S_{\!m,0}^2 \int_{-\infty}^\infty \dd u\, \Exp{(3-2\gamma)u}}_{\int_0^\infty\dd R\,R^2\,S^2_{\!m}(R)}\;\Im\Bigg\{\underbrace{\tfrac12\int_{-\infty}^\infty \dd w\, \Exp{-\imag\zeta w-\frac12|w|}\,b_{1/2}^m\Big(\Exp{-|w|}\Big)}_{K_m(\zeta)}\Bigg\},
\end{align}
\end{subequations}
which is no longer zero (not even finite), except for $\gamma=\sfrac32$, for which $\zeta$ is real and the imaginary part of the integrand of the integral over $w$ is antisymmetric and hence $\Im\{K_m(\zeta)\}=0$. Thus, when merely changing the order of integration from $\iint\dd u\,\dd u'$ to $\iint\dd u\dd (u'-u)$, the result changes. This happens because the integrand is not absolute integrable (the integral over its absolute value diverges). The situation is analogous to a series $\sum_{n=0}^\infty a_n$ that is not absolute convergent ($\sum_{n=0}^\infty|a_n|$ diverges): changing the order of summation leads to convergence to different results or even divergence. The problem arises because (i) logarithmic spirals have, for any $R>0$, an infinity of turns at radii $<R$ and at radii $>R$, and (ii) the integral over the amplitude of these power law models does not converge.

\end{appendix}
\end{document}